\documentstyle[pre,aps,epsfig,eqsecnum]{revtex}

\newcommand \rig {\right}
\newcommand \lef {\left}

\newcommand \al {\alpha}
\newcommand \bt {\beta}
\newcommand \la {\lambda}

\newcommand \lan {\langle}
\newcommand \ran {\rangle}
\newcommand \La {\Lambda}
\newcommand \de {\delta}
\renewcommand \d {{\rm d}}
\newcommand \D {{\rm D}}
\newcommand \De {\Delta}
\newcommand \pa {\partial}
\newcommand \pr {\prime}
\newcommand \tdpr {\bar t}
\newcommand \sdpr {\bar s}

\newcommand \tilt {\Gamma _J}
\newcommand \ilt {\bar{\Gamma} _J}

\newcommand \pdt {\partial _{t}}
\newcommand \pds {\partial _{s}}
\newcommand \si {\sigma}

\newcommand \seq {i_1\dots i_p}
\newcommand \lseq {1\leq i_1<\dots <i_p\leq N}
\newcommand{\BEQ}{\begin{equation}}
\newcommand{\EEQ}{\end{equation}}
\newcommand{\BEA}{\begin{eqnarray}}
\newcommand{\EEA}{\end{eqnarray}}
\renewcommand{\H}{{\cal H}}
\newcommand{\dbarrm} {\d}

\begin{document}

\draft
\title{Model glasses coupled to two different heat baths}

\author{A.E. Allahverdyan$^{1,3)}$, Th.M. Nieuwenhuizen$^{2)}$, 
D.B. Saakian$^{1)}$
}
\address{
$^{1)}${\em Yerevan Physics Institute, Alikhanian Brothers St. 2, 
Yerevan 375036, Armenia}\\
$^{2)}${\em Department of Physics and Astronomy,  $^{3)}$ 
Institute of Theoretical Physics, 
\\ University of Amsterdam, Valckenierstraat 65, 1018 XE Amsterdam, 
The Netherlands. \\ } 
E-mails: armena@wins.uva.nl, nieuwenh@wins.uva.nl, 
saakian@jerewan1.yerphi.am 
}
\date{\today}
\maketitle

\begin{abstract}
In a $p$-spin interaction spherical spin-glass model both the spins
and the couplings are allowed to change in the course of time. 
The spins are coupled to a heat bath with temperature $T$, while the
coupling constants are coupled to a bath having temperature $T_{J}$.
In an adiabatic limit (where relaxation time of the couplings is
much larger that of the spins)
we construct a generalized two-temperature thermodynamics. 
It involves entropies of the spins
and the coupling constants.
The application for spin-glass systems leads to
a standard replica theory 
with a non-vanishing number of replicas, $n=T/T_J$. 
For $p>2$ there occur at low temperatures two different glassy 
phases, depending on the value of $n$.
The obtained first-order transitions have positive
latent heat, and positive
discontinuity of the total entropy.
This is the essentially non-equilibrium effect.
The predictions of longtime dynamics and infinite-time 
statics differ only for $n<1$ and $p>2$.
For $p=2$ correlation of the disorder (leading to a non-zero $n$)
removes the known marginal stability of the spin glass phase.
If the observation time
is very large there occurs no finite-temperature spin glass
phase. 
In this case there are analogies with the 
broken-ergodicity dynamics in the usual spin-glass models
and non-equilibrium (aging) dynamics. 
A generalized fluctuation-dissipation relation is derived. 
\end{abstract}


\section{Introduction}

\noindent

The spin-glass model proposed by Edwards and Anderson \cite{parisibook,binder}
is a paradigm for large number of different random systems in
nature. Its main assumptions, randomness and quenching of exchange couplings,
reasonably reflect the crucial properties of random magnets with
localized magnetic moments
\cite{binder}, structurally disordered materials \cite{nagaev},
 and large number of other systems 
(for example, artificial neural networks \cite{parisibook}). In particular,
alternation means that there are ferromagnetic as well as 
antiferromagnetic couplings
(attractive and repulsive in the language of particle dynamics in structurally
disordered compounds and alloys \cite{nagaev}),
and quenching means that there exists a large difference between relaxation 
times of couplings and magnetic degree of freedom (spins). 

The typical example of a  spin-glass is a dilution of a magnetic 
metal (such as $Fe$ or $Mn$) in a non-magnetic host (for example 
$Cu$), where the concentration of the magnetic metal is not very large.
Thus there is no direct exchange interaction between magnetic ions, but due to
conduction electrons of $Cu$ an indirect exchange interaction is
 possible (RKKY interaction)
\cite{nagaev}. This interaction has oscillating character with respect of the 
distance between a pair of ions, and because the positions
of magnetic ions in the dilute  are random we have the typical case of
spin-glass. By a more simple  and evident mechanism spin-glass is
realized in a dilution of ferromagnetic metal with antiferromagnetic one.

However, in all these cases there can be some doubts about 
truly quenching of coupling constants. 
Furthermore, several independent mechanisms can be pointed out 
for relatively slow changing of coupling constants with time: 1) 
diffusion of magnetic ions; 
2) the distance between a pair of ions can be changed due to 
external variations of pressure \cite{UFN}.
Such reasons very naturally induce developments of spin-glass 
models where coupling constants are not quenched variables {\it a priori}, 
but change with time, according to some slow stochastic process. 
It will allow to consider the different
and perhaps unexpected  case of a large but finite separation
between relaxation times.
If we start with some typical 
state far from equilibrium, and if the total relaxation 
time of the whole system
is very large, 
then in some intermediate time regime the spins relax to 
a partial equilibrium with temperature $T$, 
while the coupling constants relax to a
partial equilibrium at some different temperature $T_J$ 
\cite{sherington,dotsenko,kondor}.
For describing 
this regime, we can formally introduce
two different thermostats
with temperatures $T$ and $T_J$.
Later we shall introduce other infinite time scales but all these times are
much smaller than the relaxation time towards the total equilibrium
with the unique temperature.
The usual spin glass model is recovered when $T_J$ is infinite, because
in this limit the coupling constants are totally independent of 
each other and of the spins.
Moreover, with these general assumptions about time-scales and
temperatures one can obtain the non-equilibrium stationary
distribution in more general cases, and construct the 
appropriate thermodynamics.

First phenomenological spin-glass models with time-dependent
coupling constants were considered by Horner \cite{horner}.
The two-temperature approach was introduced by Coolen et al. \cite{sherington}
and Dotsenko et al. \cite{dotsenko}.
Recently one of us discussed a phenomenological
two-temperature approach for describing the dynamics of model glasses
in terms of a non-equilibrium thermodynamics.
In this approach one temperature is the temperature of the thermal bath, 
while the second temperatures is self-generated and time-dependent,
due to the non-equilibrium nature of the glassy state 
\cite{theotwotem}-\cite{Nlongthermo}.
Our approach in the present paper is complementary: We assume the 
existence of two fixed temperatures for the different components. 
It can be realized that different temperatures for different degrees 
of freedom, and large difference between the corresponding relaxation 
times are two necessary
ingredients of so-called glassy behavior: Common action of these
physical mechanisms
ensures a specific, non-equilibrium regime of behavior. 

We shall investigate the two-temperature
dynamics and statics of the $p$-spin
interaction spin glass model ~\cite{crisanti1,crisanti2,theoprl1}.
Multi-spin interaction models are not only a convenient
laboratory for investigating phase transitions but also very well
known in physics of magnets 
\cite{nagaev}.
For example, this type of interaction
is necessary for understanding properties of He$_3$~\cite{fisher}
and metamagnets \cite{nagaev}.
The statics of the model is described by one step of 
Replica Symmetry Breaking (1RSB). The transition is first-order-type.
Near the spin glass transition point the order parameter has a
discontinuity \cite{crisanti1,theoprl1}, but there generally is
no latent heat. 
The dynamics of the model is analytically
tractable and has a rich structure \cite{crisanti2}.
Due to the systems sensitiveness to the interchange of the limits 
$N\to\infty$ and $t\to\infty$, there is a difference
between the long time limit of the dynamics and the statics: 
The critical temperatures for the spin glass transition which are derived
from the statics (where Gibbs distribution is assumed a priori)
and from the dynamics (Langevin equations with typical initial conditions)
are known to be different; this fact can be traced back to omission
of activated processes in the mean-field dynamics~\cite{theotwotem}.
There is a related difference in the thermodynamics
of the system derived via the two approaches \cite{theocomplexity}. 

This paper is organized as follows.
In section \ref{sec:model} 
the dynamics of the model is introduced
 via a set of Langevin equations.
In section \ref{sec:2T} we discuss the case where the coupling constants 
are fixed with respect of the dynamics of the spins (adiabatic statics). 
For this case we construct an analog of the usual thermodynamics. 
The considerations in this section are actually more
general, and do not depend on the details of the adiabatic system. 
This general theory is applied to the concrete case of the
mean-field spherical spin-glass model in section \ref{sec:stat}. 
We also discuss different phase transitions which arise
in this context. 
The detailed discussion of thermodynamical quantities is given.

In  section \ref{sec:dyn} 
the ergodic dynamic of the spin-glass model is investigated. 
As opposed to usual mean-field
spin-glass systems this dynamic has a larger field of relevance in our case. 
We show in the adiabatic case that there is a difference
between predictions of statics and long-time dynamics if the temperature of
coupling constants is not larger than the temperature of spins.  
Effects connected with 
large observation times are discussed also.
The summary of our results is represented at the last section. 
Some technical questions are considered in the appendices.

\section{The multi-spin interaction model and its heat baths}
\label{sec:model}

In the mean-field spherical p-spin model we add a harmonic energy 
for the couplings, yielding
\BEQ
\label{Htot} 
\H= \sum_{\lseq } J_{i_1\dots i_p} \sigma_{i_1}\cdots\sigma_{i_p}
+\frac{v}{2J^2_N}
\sum_{\lseq } J^2_{\seq}, 
\end{equation} 
where $J_N^2= {p!J^2}/(2N^{p-1})$ is the usual normalization
factor for mean-field models, with $J$ and $v$ being fixed energy scales.  
We assume that the coupling
constants and the spins interact with heat baths at temperatures 
$T_J $ and $T $, respectively.  The (overdamped)
Langevin equations for the dynamics in this model have the following form:  
\begin{equation} 
\label{2}
\Gamma \pdt \si _{i}=-r\si _{i}-\frac{\partial \H}{\partial \si _{i}}
 +\eta _{\,i}(t),\qquad \langle \eta _{i}(t)\eta
_{j}(t^{\prime })\rangle =2\Gamma T\delta _{ij}\delta (t-t^{\pr })
\end{equation}
\begin{eqnarray}
\label{3}
&&\ilt \pdt J_{i_1\dots i_p}
=-\frac{\partial  \H}{\partial J_{i_1\dots i_p}}
+\eta _{i_1\dots i_p}(t), \qquad \langle \eta _{i_1\dots
i_p}(t)\eta _{j_1\dots j_p}(t^{\prime })\rangle =
2\ilt T_J\delta _{i_1\dots i_p,j_1\dots j_p}\delta (t-t^{\pr })
\end{eqnarray}
Here $r(t)$ is the Lagrange multiplier for enforcing
the spherical constraint $\sum_i\si_i^2(t)=N$, 
while  $\Gamma$ and $\ilt $ are the damping constants.
The coupling constants $J_{i_1...i_p}$ and the noises $\eta _{\seq }$ are
symmetric with respect of interchange of the indices. 
In Eqs. (\ref{2},\ref{3}) so called Einstein
relation holds between the strength of noise and the damping constant. 
This means that the thermal
baths themselves are in thermal equilibrium \cite{gardiner}.
For ensuring the correct 
thermodynamical limit we must
take: $\ilt =\tilt / J^2_N$ 
(see Appendices A, B). 

Straightforward calculations show that if $v\sim T_J$  and
the limit $\tilt \mapsto \infty$ is taken first, 
followed by $T_J \mapsto \infty $, then
the coupling constants are quenched (with respect to the spins) 
independent Gaussian random variables. This limit thus yields the standard
$p$-spin model (see Appendix A).

With help of standard methods \cite{thirumalai,dynfunc} we 
study the dynamics in Appendix B. 
We arrive at the following equations for the average dynamics of a 
single spin in the mean field caused by the other ones
\begin{eqnarray}
\label{4}
&  &(\Gamma \pdt +r)\si (t)=\frac{pJ^2}{2\Gamma _J}
\int_{-\infty}^{t} d\tdpr e^{-(t-\tdpr )/ \tau _J } 
C^{p-1}(\tdpr ,t)\si (\tdpr
)\nonumber \\
&  &+\frac{pT_J J^2}{2v} (p-1) 
\int_{-\infty}^{t} d\tdpr e^{- (t-\tdpr )/\tau _J }
C^{p-2}(\tdpr ,t)G(t,\tdpr )
\si (\tdpr ) +\eta (t), \nonumber \\
& &
\langle \eta (t)\eta (t^{\prime })\rangle =2\Gamma T \delta (t-t^{\pr
})+\frac{pT_J J^2}{2v}
\exp{(-|t-t^{\pr }|/\tau _J)} C^{p-1}(t,t^{\pr }),
\end{eqnarray}
where
\begin{equation}\label{4a}{\tau _J} =\frac{\tilt }{v}\end{equation}
is the timescale at which  the couplings change.
Details of the derivation of this equation can be found in the appendix B.

Some comments about the general structure of Eq. (\ref{4}) are at order.
As it is well-known that the effective dynamical equations for 
spin-glass systems with quenched disorder are 
essentially non-markovian, i.e., they depend on
the "history" of the process. Evidently, this
arises due to quenching  of a coupling constant with time. 
In our case | on account of the 
characteristic time (\ref{4a})| the non-markovian property is 
"smoothened" by the exponential kernel (see Eq. (\ref{4})).

With help of (\ref{4}) we may derive coupled equations for
the correlation function
\begin{equation}
\label{d1}
C(t,t ^{\prime })=
\frac{1}{N}\sum_{i}\left \langle \si _i(t)\si _i(t^{\pr })\right \rangle
\end{equation}
and the response function
\begin{equation}
\label{d2}
G(t,t^{\pr })=\frac{1}{N}\sum_{i}
\frac{\delta \left\langle\si _i(t)\right\rangle}{\delta  h _i(t^{\pr })}
\end{equation}
describing the response of spin $\sigma_i$ to a small
local field $h_i$ imposed on an  earlier moment $t'$, via an instantaneous
 change of the Hamiltonian as $\H\mapsto \H-h_i\si _i$.
To fix the units, we shall take $\Gamma =1$
from now on. We find for $t>t'$
\begin{eqnarray}
\label{5}
&  &(\pdt +r)C(t,t^{\pr })=\frac{pJ^2}{2\Gamma _J} \int_{-\infty}^{t}
d\tdpr e^{-(t-\tdpr )/\tau _J} C^{p-1}(\tdpr ,t)C(\tdpr ,t^{\pr })
\nonumber \\ &
&+\frac{pT_J J^2}{2v} (p-1) \int_{-\infty}^{t} 
d\tdpr e^{-(t-\tdpr )/\tau _J} C^{p-2}(t
,\tdpr)G(t,\tdpr ) C(\tdpr ,t^{\pr })\nonumber \\
&&+\int_{-\infty}^{t^{\pr }} d\tdpr G(t^{\pr },\tdpr ) \left (2 T \delta
(t-\tdpr)
+\frac{pT_J J^2}{2v} e^{-|t-\tdpr |/\tau _J}C^{p-1}(t,\tdpr )\right ) 
\end{eqnarray}
\begin{eqnarray}
\label{6}
&  &(\pdt +r)G(t,t^{\pr })=\frac{pJ^2}{2\Gamma _J} \int_{-\infty}^{t}
d\tdpr e^{-(t-\tdpr )/\tau _J} C^{p-1}(\tdpr ,t)G(\tdpr ,t^{\pr })
\nonumber \\
& &+\frac{pT_J J^2}{2v} (p-1) \int_{-\infty}^{t} 
d\tdpr e^{- (t-\tdpr )/\tau _J}
C^{p-2}(t
,\tdpr)G(t,\tdpr ) G(\tdpr ,t^{\pr }).
 \end{eqnarray}
Generally speaking, both the relaxation toward a stationary state as well as 
fluctuations in that state are described by this closed pair of
equations. Particularly,
in the second case the time-translational invariance is expected to
hold: one-time quantities
do not depend of time, two-time quantities depend only on the
difference of times:
\begin{equation}
\label{d3}
C(t,t^{\pr })=C(t-t^{\pr }),\  \
G(t,t^{\pr })=G(t-t^{\pr }).
\end{equation}
This regime only applies in the limit when the initial time $t_0$
goes to $-\infty$; this was already inserted in the lower
limits of integration of Eq. (\ref{6}).
Indeed, then the memory of the initial conditions is washed out, 
and the system relaxes toward 
its stationary state.
It should be stressed that this infinity is taken only 
after the thermodynamic limit $N\to \infty$.
$t-t^{\pr }$ can be viewed as an observation time, 
or as the finding of some clock designed 
to display the temporal dynamics of fluctuations.

\section{Two-temperature adiabatic thermodynamics: general structure }
\label{sec:2T}
\  $  $ \
Recently Coolen, Penney, and Sherrington \cite{sherington} proposed
a dynamical approach to the statistical mechanics of spin glass systems,
 where the introduction of replicas is not needed initially (though
they enter later without the $n\to 0$ limit; see also 
\cite{landauer}-\cite{dotsenko}).
This approach can be called adiabatically static,
because it is a static limit obtained by  taking
\begin{equation}
\label{pun1}
\tilt \mapsto \infty
\end{equation}
(i.e., $\tau _J \mapsto \infty $ but $v$ remains finite)
immediately in the initial equations of motion (recall that $\tau _J $ 
remains much smaller 
than the initial time: $\tau _J \ll |\,t_0|$).
For times much less than $\tau_J$ the spins will still see random couplings.
As opposed to the standard case, the couplings are no longer uncorrelated.
The correlation of couplings is coded in a finite $T_J$, and may lead
to new physics. 

Equations (\ref{2}),(\ref{3}) can be investigated by the method
of adiabatic elimination (see for example \cite{gardiner}). 
Here we go further and construct the corresponding thermodynamics.
For the case of a static distribution the procedure is as follows.
 First equation (\ref{2}) is solved keeping the coupling constants
$J$ fixed (adiabatic following). Further, the Langevin equations in 
this case have the following equilibrium distribution 
\begin{equation}
\label{pun2}
P(\sigma |J)=\frac{1}{Z_\si (J)}\exp   [-\beta \H (\si ,J) ]
\end{equation}
(In this section we do not write the spherical factor explicitly; 
the reader can consider it to be included in $ \H (\si ,J)$.)
The partition sum for given $J$-configuration is
\begin{equation}
\label{pun3}
Z_\si(J)={\rm Tr}_{\si }\exp \left  [- \beta \H(\si ,J) \right ]
\end{equation}
In the evolution of the $J$-subsystem the averaging over
 the fast variables in (\ref{3}) can be carried out.
In general this should be in a self-consistent way.
At quasi-equilibrium of the $\si$-subsystem
this average can be performed and leads to the use of (\ref{pun2}).
In this way we get from eqs. (\ref{Htot}), (\ref{2}) and (\ref{3}) 
a related dynamics for the couplings, in which $\H (\si , J)$
is replaced by $-T\ln Z_\si (J)$, 
which plays the role of effective hamiltonian in the corresponding dynamics. 
Specifically we have
the effective equation of motion
\begin{equation}
\label{pun4}
\tilt \pdt J_{i_1\dots i_p}
=\partial _{J_{i_1\dots i_p}}
T\ln Z_\si(J)
+\eta_{i_1\dots i_p}(t)
\end{equation}
As the noise is due to a bath at temperature $T_J$, see eq. (\ref{3}), 
the equilibrium distribution of this equation reads
\begin{equation}
\label{pun5}
P(J)=\frac{Z^n_\si(J)}{{\cal Z}}
\end{equation}
where
\begin{equation}
\label{pun55}
{\cal Z}=\int DJZ^n_\si(J)
\end{equation}
This approach introduces a  `dynamical' replica index
\begin{equation}
n=\frac{T}{T_J}
\end{equation}
The free energy follows from the usual formula
\begin{equation}
\label{pun6}
F=-\frac{T}{n}\ln {\cal Z}=-T_J\ln {\cal Z}
\end{equation}
Now we construct the appropriate thermodynamics for this
two-temperature model using eqs. (\ref{pun5}), (\ref{pun55}).
We first need the joint distribution of $\si $ and $J$. As usual  we can
express the unrestricted probability $P(\si,J)$ in terms of
the conditional probability $P(\sigma |J)$  as
\begin{equation}
\label{pun7}
P(\sigma ,J)=P(J)P(\si |J)=\frac{Z^{n-1}_{\si }}{{\cal Z}}\exp(-\bt \H (\si,J))
\end{equation}
For the total energy $U$
we have
\begin{equation}
\label{pun8}
U={\rm Tr}_{\sigma }\int \D J\,\, \H (\si ,J)P(\si ,J),
\end{equation}
which can be viewed also as the average energy of the $\si $-subsystem.
Direct calculations show that
\begin{equation}
\label{pun9}
-\frac{1}{n}\frac{\partial }{\partial \bt}\ln {\cal Z}\Bigl|_n\Bigr.=U, \ \
-\frac{1}{n}\frac{\partial }{\partial \bt }
\ln {\cal Z}\Bigl|_{T_J}\Bigr.=U-F_J, 
\end{equation}
where
$$
F_J=-\int \D J\,\, T\ln Z_\si(J)P(J)
$$
is the self-averaged free energy of the $\si $-subsystem
or the mean effective energy. 

We  define entropies of total system and its subsystems by the usual
Boltzmann-Gibbs-Shannon formula with help of the corresponding
distributions (\ref{pun7}). For example, the total entropy reads
\begin{equation}
\label{pun12}
S=-\int \D J\, {\rm Tr}_{\sigma }\, P(\si ,J)\ln P(\si ,J),\  \
\end{equation}
This involves just the general, statistical definition of 
entropy for a macroscopic
system,  which is also relevant outside equilibrium \cite{strat}.
Due to the decomposition (\ref{pun7}) one gets two contributions,
\begin{equation}
\label{pun15a}
 S=S_{\si}+S_J
\end{equation}
where
\begin{equation}
\label{pun15b}
S_\si=\int \D J\, P(J)\left\{-{\rm Tr}_{\sigma }\, P(\si|J)\ln P(\si|J)\right\}
\end{equation}
is our analog of the usual quenched average entropy of the spin
motion (more precisely, it is the so-called 
conditional entropy \cite{strat}), while the coupling-part of the
entropy reads
\begin{equation}\label{pun15c}
S_J=-\int \D J \,\,P(J)\ln P(J)
\end{equation}
The analogous separation of the total entropy in two or more
 parts appears in other problems of statistical physics \cite{ben}
\cite{werner}. In particular, it concerns the fine-graining 
procedure which introduces states of a statistical system relative to a
fixed value of some properly chosen order parameter 
(a quantity under macroscopic control). 
Then the total entropy is also separated 
as in Eq. (\ref{pun15a}), where $S_{\si}$ corresponds to average
 entropy of the relative states, and $S_J$ corresponds to entropy of the
order parameter itself. A simple example is just a piston 
separating a volume with a gas in two equivalent parts. Then $S_{\si}$ is 
entropy of the gas in one part, and obviously $S_J=\ln 2$. 
A similar separation occurs in glassy transitions, where $S_J$ 
corresponds to the configrational entropy or complexity
~ \cite{theotwotem}\cite{theocomplexity}.
The famous phenomena of Maxwell's demon \cite{ben} 
(proposed by Maxwell more than one century ago),
and its subsequent reformulations
and generalizations display that in a process of measurement 
$S_{\si}$ can be decreased, instead $S_J$ increases, and the total entropy
$S_{\si}+S_J$ can only increase.

Let us consider now a generalized thermodynamics which arises 
in the context of Eqs. (\ref{pun6})-(\ref{pun15c}).
As a matter of fact, to generalize the usual thermodynamics we 
notice the following relations
only
\begin{equation}
\label{pun13}
F=F_J-T_JS_J
\end{equation}
\begin{equation}
\label{terra1}
F=U-T_JS_J-TS_{\si}
\end{equation}
Eq. (\ref{pun13}) is the usual thermodynamical formula for the $J$-subsystem.
The second formula is more interesting because
it gives a connection between the characteristics of the subsystems
and the whole system.
This agrees with the expression of the free energy for a glassy system
put forward  previously by one of us~\cite{theotwotem,NEhren,Nhammer}.
In that approach the equivalent of $T_J$ is the dynamically generated
effective temperature. 

These results can be written in the differential form
\begin{equation}
\label{3.36}
\d F=-S_{\si}\, \d T -S_J \,\d T_J
\end{equation}
If we add an external field, it can be checked  that
 \begin{equation}\label{dF=}
\d F=-S_{\si}\, \d T -S_J \,\d T_J-M\,\d H
\end{equation}
This implies that the first law of thermodynamics takes the form
 \begin{equation}
 \label{3.38}
\d U=T \,\d S_{\si}+T_J\,\d S_J -M\,\d H
\end{equation}
As the last term can be indentified with $\dbarrm W$,
the work done on the system,  the change in heat reads
\BEQ \label{dQ=}\dbarrm Q=T \,\d S_{\si}+T_J\,\d S_J \EEQ
Eqs. (\ref{3.36}, \ref{dF=}, \ref{3.38}) constitute a 
manifestation of a thermodynamic process in the following sense. 
A process is called ``thermodynamic''
if its characteristic time $\tau _{th}$ is much larger than 
internal relaxation times of the considered system. 
Due to this condition it is possible to represent
the process as a chain of stationary states. 
In our adiabatic system there are
two relaxation times, $\tau $ 
(an effective characteristic time of spins) and $\tau _J$ with 
$\tau _J \gg \tau $. There can thus exist two types of thermodynamic
processes: A slow one with
$\tau _{th}\gg \tau _J \gg \tau  $ and a relatively 
fast one with $\tau _J \gg
\tau _{th}\gg  \tau $. In the second case
the $J$-subsystem does not change during this process,
implying, for example, $\d S_J=0$. We see that
(\ref{3.36}-\ref{3.38}) represent
a slow thermodynamic process where states of both subsystems 
are changed. This classification allows us to discuss
the question about heating or cooling of a two-temperature adiabatic system. 
If cooling is slow enough, typically both temperatures change; for example,
in the extreme cooling process we can have $T_J\to 0$ and $T\to 0$
simultaneously. In the opposite case of a fast cooling (or heating) process 
$T_J$ is a constant while $T$ varies in time.

Irreversible effects also can be included in the present scheme. 
In general, irreversibility means that there are additional
 sources to increase entropy or decrease free energy. Namely it reads:
\begin{eqnarray}
\label{kenguru}
&&\d F<-S_{\si}\, \d T -S_J \,\d T_J-M\,\d H, \nonumber \\
&&\d U<T \,\d S_{\si}+T_J\,\d S_J -M\,\d H
\end{eqnarray}

Eq. (\ref{terra1}) can be easily generalized to a many-level
adiabatic system where the first part of variables is
slow with respect to the second part, the second part is slow with respect to
third part,...
For example, if we have a three-level system with the parts:
 $\{ J\} $, $\{ \si_1\} $, $\{ \si _2\} $, having relaxation times
$\tau _J\gg \tau _{\si _1}\gg \tau _{\si _2}$
\begin{equation}
\label{terra2}
F=U-T_JS_J-T_{\si _1}S_{\si _1}-T_{\si _2 }S_{\si _2}
\end{equation}
It should be noticed, however, that on time scales of order $\tau _{\si_1}$,
where interesting non-equilibrium dynamics of the $\si_1$ system occurs,
the $\si_2$-system is in equilibrium, while the $J$-system is fixed;
on the other hand, on timescales of order $\tau_J$ both the $\si_1$
and the $\si_2$ systems are in equilibrium. A physical realization of
this scenario occurs in  glass forming liquids, with their
 fast and slow $\beta$ processes, while
the $J$-system then describes the configurational or $\alpha$-processes.
In this context Eq. (\ref{terra2}) corresponds also to recently introduced
models with two-level disorder: Coupling constants are also considered
as  frozen variables with respect to some other set 
of variables \cite{sherington2}\cite{sherington3}.

\section{Adiabatic statics.}
\label{sec:stat}
\ $     $ \
In this section we investigate the adiabatic static limit of the mean field
spherical spin-glass model introduced in the
section \ref{sec:model}. The free energy is described by (\ref{pun6}).
As in ref \cite{crisanti1} we have
\begin{equation}
\label{a1}
{\cal Z}=
\prod _{\al \bt }\left (
\int \frac{dq_{\al \bt }d\la _{\al \bt }}{2\pi i}
\right )
\exp{(-NG_n(q_{\al \bt },\la _{\al \bt }))}
\end{equation}
\begin{eqnarray}
\label{a2}
2G_n(q_{\al \bt }\la _{\al \bt })
&=&-n\ln 2\pi -\frac{\mu }{p}\sum_{\al \bt }q_{\al \bt }^{p}
+\sum_{\al \bt }q_{\al \bt }\la _{\al \bt }
+\ln {\rm det}(-\la ) \nonumber\\
&=& {\rm const.} -\frac{\mu }{p}\sum_{\al \bt }q_{\al \bt }^{p}
\,\,-\,{\rm tr}\,\ln(q ),
\end{eqnarray}
where $\mu =pT_J J^2/2vT^2$,
$q_{\al \bt }=\langle \si _{\al }\si _{\bt }\rangle $ is the usual order
parameter describing the spin-glass ordering,
and $\la _{\al \bt }$ are Lagrange multipliers \cite{parisibook,binder}.
\footnote{Besides (\ref{a2}) there is a contribution to the free 
energy which arises from the integration by the coupling constants, 
and has an order ${\cal O} (N^p\ln N)$.
Usually this contribution is omitted (see \cite{dotsenko} for example),
because it does not depend on the order parameter. 
An alternative point of view is to consider (\ref{a2}) as the leading 
finite-size effect.}
We have three independent parameters: $T$, $T_J=T/n$, and $v$.
In this paper we consider phase transitions only in the following subspace of
three dimensional space of the parameters: $v= T_J$, $n=T/T_J$ is
fixed, and the relevant parameter is $T$ 
(of course, all such regimes with $v\sim T_J$ are qualitatively equivalent). 
Thus for $\mu $ we have the standard expression
\begin{equation}
\label{mu}
\mu=\frac{p\beta ^2J^2}{2} 
\end{equation}
As we have discussed in the previous section, there is some regime of cooling
where $n$ indeed can be a constant. 

\subsection{Replica symmetric solution}
\label{subsec:rs}
\  $  $ \
In the investigation of a spin-glass the first step 
is to make the Replica Symmetry (RS)
assumption for the order parameter \cite{parisibook,binder}:
$q_{\alpha \beta}=q$ (for $\alpha \not= \beta$),
where $q$ is the usual Edwards-Anderson parameter. 
In other words, one assumes that there is 
only one thermodynamical state (up to possible global symmetry 
transformations).
The expression for the RS free energy  $f_{rs}=F_{rs}/N$ has
the following form
\begin{equation}
\label{a3}
2\bt f_{rs}=-\ln (1-q)-\frac{1}{n}\ln \left (
1+\frac{nq}{1-q}
\right )-\frac{\mu }{p}(1+(n-1)q^p)
\end{equation}
where $q$ is determined by the saddle point equation
\begin{equation}
\label{a4}
\mu q^{p-1}=\frac{q}{(1-q)(1+q(n-1))}
\end{equation}
In section  \ref{sec:dyn} we shall see that this equation can be
obtained from the long-time statics if a slow dynamics for the
coupling constants is assumed.
Following (\ref{pun12})-(\ref{terra1}) we get for energy and entropies
\begin{equation}
\label{kamo11}
2\bt u_{rs}=-\frac{\mu
}{p}(1+(n-1)q^p)
\end{equation}
\begin{equation}
\label{kamo22}
2 s_{\sigma }=\ln (1-q)
+\frac{q}{1-q+nq}
-\frac{\mu}{p}(1-q^p)
\end{equation}
\begin{equation}
\label{kamo33}
2 s_{J}=
\ln (1+\frac{nq}{1-q})
-\frac{nq}{1-q+nq}
+\frac{n\mu}{p}(1-q^p)
\end{equation}

It is well known that from a point of view of phase transitions the set of
$p$-spin models can be divided into the two main groups: $p>2$ and $p=2$. 
In the first case the qualitative phase diagram of
the model does not depend on $p$ (as long as it is finite).
Thus in the main part of this subsection we investigate the case $p>2$;
 our results for $p=2$ will be presented at the end.

We start with investigation of Eq. (\ref{a4}). For high temperatures 
there is only the  paramagnetic phase with $q=0$. 
The critical point $T_{1,rs}$ can be defined as the first temperature
where a non-zero solution of (\ref{a4}) is possible. 
For $p>2$ there occurs a first-order phase transition point
with discontinuity of the order parameter
\begin{equation}
\label{serso1}
q_{1,rs}=\frac{\sqrt{(2-n)^2(p-1)^2+4p(p-2)(n-1)}+(n-2)(p-1)}{2p(n-1)}
\end{equation}
The related onset temperature is 
\begin{equation}
\label{T1rs}
T_{1,rs}=J\, \sqrt{\frac{p}{2}\, q^{p-2}_{1,rs}\, (1-q_{1,rs})\, (1+(n-1)q_{1,rs})}
\end{equation}
Some limiting cases can be investigated; 
for example, if $p$ is large enough we get
\begin{equation}
\label{plarge}
q_{1,rs}\sim 1-\frac{1}{p}, \  \  T_{1,rs}\sim \sqrt{\frac{n}{2e}}
\end{equation}
Any physical solution must be stable against small perturbations, 
therefore the analysis of linear stability
for a possible replica-symmetric solutions should be perfomed.
The eigenvalues of the corresponding Hessian for a finite $n$ were computed in
\cite{crisanti1}.
There are three main sectors of fluctuations and the three corresponding
eigenvalues:
\begin{equation}
\label{a5}
\La _1=-\mu (p-1)q^{p-2}+\frac{1}{(1-q)^2}, \  \
\sum_{\bt }\delta q_{\al \bt }=0
\end{equation}
\begin{equation}
\label{a55}
\La _2=\La _1-\frac{(n-2)q}{(1-q)^2(1+(n-1)q)}, \  \
\sum_{\bt }\delta q_{\al \bt }\not = 0, \  \ \sum_{\al \bt }\delta q_{\al \bt }
= 0
\end{equation}
\begin{equation}
\label{a6}
\La _3=\La _1-
\frac{(n-1)q}{(1-q)^2(1+(n-1)q)}
\left (2-\frac{nq}{1+(n-1)q}\right ),
\ \  \sum_{\al  \bt }\delta q_{\al \bt }\not =0
\end{equation}
The first eigenvalue, the so-called ``replicon'' or ``ergodon'',  
is displayed by the most coherent fluctuations 
(RSB is checked usually
by this eigenvalue); the third
eigenvalue corresponds to the most non-coherent fluctuations,
and the second one takes an intermediate position. 
Before investigating  the stability of the
non-zero RS solution, 
we first discuss which eigenvalue is relevant for different values 
of $n$, so which is the smallest one.
A simple analysis shows that for $n<1$ $\La _1$  is relevant,
while for $n>1$ $\La _3$ is the most dangerous one.
This result is important: as we have seen, the relevant sector of 
the fluctuations depends on $n$, implying that the
whole structure of the phase space has strong  dependence on $n$ also.
At $n=1$ $\La_2$ becomes relevant too, but it causes no
extra problem, since we have $\La _1=\La _2$.

For $p>2$ the paramagnetic solution $q=0$ is stable everywhere (the case $p=2$
will be discussed at the end of this section). Now we check the stability of
the nonzero solution of Eq. (\ref{a4}) for $n<1$ .
This solution monotonically decreases from (\ref{serso1}) to zero.
 From positivity of $\La _1$ we get
\begin{equation}
\label{rsstab}
q\geq \frac{p-2}{n+p-2}
\end{equation}
Thus the nonzero solution is stable only for
\begin{equation}
\label{rsstab1}
T\leq T_{rs,st}=J\, \sqrt{ \frac{n^2p(p-1)(p-2)^{p-2}}{2(n+p-2)^p}}
\end{equation}
Because $T_{rs,st} <T_{1,rs} $ we see that the solution is stable only for
sufficiently low
temperatures. In particular, for $n=0$ it is unstable for every
temperature, and that explains why this solution could be discarded till now.
On the other hand, we shall show that if
$n\geq 1$ the  RS solution is stable everywhere.
Indeed, for the nonzero solution of (\ref{a4}) $\La _3$ is positive, and has
the following form:
\begin{equation}
\label{serso5}
\La _3=\frac{p(n-1)q^2+(2-n)(p-1)q+2-p}{(1-q)^2(1+(n-1)q)^2}
\end{equation}
Further discussion about properties of replica symmetric spin-glass solution 
will be given after consideration 
of replica symmetry breaking solutions.
In particular, we shall see that there are temperatures ($T_{2,rs}$, $T_{3,rs}$) 
of the true thermodynamical phase transitions.

\vbox{
\begin{table}[bhb]
{\footnotesize
$$
\begin{array}{||c|c|c||}
\hline n & T_{1,rs} & T_{2,rs}   \\ \hline
 1.01       &0.6125      &  0.5862         \\ \hline
 1.10       &0.6271      &  0.6001         \\ \hline
 2.50       &0.8270      &  0.7543         \\ \hline
 4.00       &1.0052      &  0.9213         \\ \hline
 6.00       &1.2049      &  1.1140         \\ \hline
 10.0       &1.5289      &  1.3567         \\ \hline
 \end{array}
$$
}
\caption{ 
 $T_{1,rs}$, and $T_{2,rs}$ for different
$n$ ($n>1$), and $p=3$. 
At $T=T_{1,rs}$ the RS spin-glass phase first appears as 
metastable one. The true first-order phase transition 
from the paramagnetic phase occurs at $T=T_{2,rs}$ (J=1).
}
\end{table}
}
\subsubsection{The case $p=2$}
\label{p=2}
Let us now consider the special case $p=2$.
It is well known that the model with $n=0$ is described by a replica symmetric
Anzatz and has interesting properties \cite{CuPa}.
For example, the $\La _1$ eigenvalue
vanishes everywhere in the low temperature spin-glass phase.
For our solution  we see (from (\ref{a4})) that in the same model
$\La_1$ is always positive for nonzero $n$, since it holds that
 \begin{equation}
 \label{serso7}
  \La _1=\frac{nq}{(1-q)^2(1+(n-1)q)}>0
  \end{equation}
So the correlation between the couplings
removes the zero modes and stabilizes the structure of the vacuum.
Since $\La _3(p=2, n>1)$ is also a positive function we see that
for $p=2$ the RS solution is stable for all $n$, and for all $T$.
It will be evident later that a consistent RSB solution is impossible 
in this case. In other words, the model with $p=2$ is completely described
by the RS anzatz.

We get for $p=2$
 \begin{equation}
 \label{p1}
 q_{1,rs}=\frac{(|n-2|+n-2)}{4(n-1)}
 \end{equation}
Now if $n\leq 2$ the phase transition is second-order, and $T_{1,rs}$ 
coincide with true phase
 transition point into the spin-glass phase. For $n>2$ the phase 
 transition is
first-order with the following scenario: 
besides $T_{1,rs}$, where the spin-glass phase occurs at the first time, 
there is also the true phase transition point $T_{2,rs} < T_{1,rs}$, 
that is determined by comparing free energies
of paramagnet and spin-glass. These statements 
can be illustrated analytically in the case of positive but small $n-2$. 
Starting 
from Eq. (\ref{a3}) we get:
\begin{equation}
 \label{puk1}
 q_{1,rs}=\frac{n-2}{2}, \  \  q_{2,rs}=\frac{2(n-2)}{3},
 \end{equation}
\begin{equation}
 \label{puk2}
 T_{1,rs}=J\left ( 1+\frac{(n-2)^2}{8}\right ), \  \  
T_{2,rs}=J\left ( 1+\frac{(n-2)^2}{9}\right ).
 \end{equation}
The difference between free energies of the spin-glass phase
 and paramagnet is positive at the transition point $T=T_{1,rs}$:
\begin{equation}
 \label{puk3}
\Delta f(T_{1,rs})=J\frac{(n-2)^4}{384}>0.
 \end{equation}
It means that the spin-glass phase appears as metastable one 
for $n>2$. The free energies are equal at the second transition
 point: $\Delta f(T_{2,rs})=0$.  Therefore this point must
be considered as the temperature of the true phase transition from
 paramagnet to spin-glass. Indeed, for $T<T_{2,rs}$ we have
 $\Delta f(T)<0$ as it should
be, since the spin-glass phase is more stable. Such a type of
 phase transition will be extensively discussed later.

We have seen that a finite $n$ removes the marginal states for 
the $p=2$ model. In the section \ref{sec:dyn} we shall see
that a similar statement holds also for the case $p>2$. 
But then the condition $n>1$ is needed for stabilizing the corresponding
marginal states.

\subsection{Replica symmetry breaking.}
\label{subsec:rsb1}
\  $ $ \
Now we investigate Replica Symmetry Breaking (RSB) solutions.
As we have seen in the previous subsection, for $n>1$ 
the stability of a RS spin-glass solution gives some hint about irrelevance
of RSB in this range of $n$.
We shall, however, first discuss the case $n<1$.
We consider here only the first step of RSB (1RSB) because a
more general type of RSB is not possible in this model; 
we omit the proof 
because it can be found in \cite{crisanti1}.

Taking the usual steps \cite{parisibook} \cite{binder} we get the following equations
\begin{eqnarray}
\label{dopdop1}
&& 2\bt f_{rsb}=-\frac{m-1}{m}\ln (1-q_1)
-\frac{n-m}{nm}\ln (1-(1-m)q_1-mq_0)\nonumber \\
&& -\frac{1}{n}\ln (1-(1-m)q_1+(n-m)q_0)-\frac{\mu
}{p}(1-q^p_1+m(q^p_1-q^p_0)+nq^p_0)
\end{eqnarray}
where $q_{\al \bt}$ takes the values $q_1$ and $q_0$, and
$m$ is the RSB parameter. 
For $n=0$ we recover the usual 1RSB equations \cite{crisanti1}.
Following (\ref{pun12})-(\ref{terra1}) we get for energy and entropies
\begin{equation}
\label{kamo1}
2\bt u_{rsb}=-\frac{\mu
}{p}(1-q^p_1+m(q^p_1-q^p_0)+nq^p_0)
\end{equation}
\begin{eqnarray}
\label{kamo2}
&&2 s_{\sigma }=\frac{m-1}{m}\ln (1-q_1)
+\frac{1}{m}\ln (1-(1-m)q_1-mq_0)\nonumber \\
&&+\frac{q_0}{1-(1-m)q_1+(n-m)q_0}
-\frac{\mu}{p}(1-q^p_1+m(q^p_1-q^p_0))
\end{eqnarray}
\begin{eqnarray}
\label{kamo3}
&&2 s_{J}=
-\ln (1-(1-m)q_1-mq_0)+\ln (1-(1-m)q_1+(n-m)q_0)\nonumber \\
&&-\frac{nq_0}{1-(1-m)q_1+(n-m)q_0}
+\frac{n\mu}{p}(1-q^p_1+m(q^p_1-q^p_0))
\end{eqnarray}
The saddle  point equations are
\begin{equation}
\label{d7}
\mu (q_1^{p-1}-q_0^{p-1})=\frac{ q_1-q_0}{(1-q_1)(1-(1-m)q_1-mq_0)}
\end{equation}
\begin{equation}
\label{d8}
\mu q_0^{p-1}=\frac{q_0}{(1-(1-m)q_1-mq_0)(1-(1-m)q_1-(m-n)q_0)}
\end{equation}
On time scale $\tau_J$ the spin system will, by definition, be in equilibrium.
Therefore $m$ is determined also by its saddle point equation,
\begin{eqnarray}
\label{a10}
& &\frac{\mu }{p} (q_1^{p}-q_0^{p})=
\frac{1}{m^2}\ln \left (1+\frac{m(q_1-q_0)}{1-q_1}\right )\nonumber \\
& &-\frac{n-m}{nm}\frac{q_1-q_0}{(1-(1-m)q_1-m q_0)}
-\frac{1}{n}\frac{q_1-q_0}{(1-(1-m)q_1-(m-n)q_0)}
\end{eqnarray}
There is some other possibility for fixing $m$: 
As was shown before \cite{thirumalai}
\cite{crisanti2} \cite{theoprl1}
this parameter can be fixed also by the
so called "marginality condition". The resulting theory describes metastable
states, in a manner also monitored by the dynamics
\cite{crisanti2} (see also section \ref{sec:dyn}). In this paper we consider
only the purely static condition (\ref{a10}).

First we note that  there is a remarkable analogy between
the RS finite-$n$ free energy (\ref{a3})
and the free energy of the 1RSB solution with $q_0=0$.
If we  interchange $q_1$ with $q$
and $m$ with $n$ we arrive at identical expressions.
This analogy  between corresponding free energies was considered in the SK
model with similar but more complicated techniques including 
two types of frozen variables \cite{sherington2}. 
Recently XY spin-glass model has been investigated by the same 
approach \cite{sherington3}.
We shall show that a similar, perhaps more informative  
analogy can be found in the dynamics of the present model.

As  well-known, the physical interpretation of replica symmetry 
breaking is connected with decomposition of the phase space
into pure states (ergodic components) \cite{parisibook}
\cite{binder}. This structure is contained 
in the overlap $q_{\alpha \beta}$.
In particular, for 1RSB the values $q_1$, $q_0$ can be interpretated 
as self-overlap of a pure state and mutual overlap between two pure states.
This information is coded also in the probability distribution of overlaps:
\begin{equation}
\label{prem}
P(q)=\frac{1}{n(n-1)}\sum_{\alpha \not = \beta}\delta 
(q-q_{\alpha \beta})=\frac{1-m}{1-n} \delta (q-q_{1})
+\frac{m-n}{1-n}\delta (q-q_{0})
\end{equation}
It is assosiated with the fraction of matrix elements 
$q_{\alpha \beta}$ which take the value $q_1$ or $q_0$ \cite{parisibook}.
We see that for interpretation of
 $(1-m)/(1-n)$ and $(m-n)/(1-n)$ as (non-negative !) probabilities we need:
\begin{eqnarray}
\label{m}
&&n<m<1 \  \ {\rm for}\  \ n<1 \nonumber \\
&&1<m<n \  \ {\rm for}\  \ n>1.
\end{eqnarray}
One could expect that the conditions (\ref{m}) are satisfied 
automatically if other more obvious physical conditions
(for example, $q_1>q_0$)
are valid. However, it is not so. Later we shall show that they
should be considered as additional conditions selecting the correct solution.

\subsubsection{Replica symmetry breaking with vanishing lower plateau}
\label{vanish}

Let us now discuss the solution of Eqs. (\ref{d7}), (\ref{a10}) 
for the case $q_0=0$.
Then the solution itself becomes independent on $n$. However, the
 dependence on $n$ does enter through Eqs. (\ref{prem}), (\ref{m}).
The considered solution has partly been investigated in \cite{crisanti1}. 
First we note that
there is a convenient parametrisation
of Eqs. (\ref{d7}), (\ref{a10})  \cite{crisanti1,parisi}. If we denote
\begin{equation}
\label{xwz1}
c=\frac{mq_1}{1-q_1}
\end{equation}
then for this quantity we get the temperature independent equation
\begin{equation}
\label{xwz2}
\frac{c^2}{p}=(1+c)\ln (1+c)-c
\end{equation}
The positive solution of this equation should be selected.
(The authors of \cite{crisanti1} employ the
slightly different variable $y=1/(c+1)$).
Taking this into account, the equation for $q_1$ reads
\begin{equation}
\label{ogul3}
\mu q_1^{p-1}=\frac{q_1}{(1+c)(1-q_1)^2}
\end{equation}
The highest temperature for which this equation has a non-zero 
solution will be denoted by $T_{1,rsb}$:
\begin{equation}
\label{ogul44}
T_{1,rsb}=J\sqrt{\frac{2(1+c)}{p} \left (\frac{ p-2}{ p}\right) ^{p-2} },
\end{equation}
and $q_1$ at this point has the value
\begin{equation}
\label{ogul4}
q_1(T_{1,rsb})=\frac{ p-2}{p}.
\end{equation}
Further, using (\ref{xwz1}) we get
\begin{equation}
\label{ogul5}
m(T_{1,rsb})=\frac{2c}{p-2}.
\end{equation}
This value is greater than $1$ for all $p>2$. The parameter $m$ 
monotonically decreases with temperature from (\ref{ogul5}) at
 $T=T_{1,rsb}$ to zero at $T=0$
(see (\ref{xwz9})).
It means that for $n>1$ only a part of the solution from 
$m(T_{1,rsb})$ to $m=1$ can be physically permissible; 
otherwise we get physically meaningless results for 
the weights (\ref{prem}). Namely, if $n$ is in the interval
 $1<n<m(T_{1,rsb})$,  then 
the temperature where the solution appears as physical one 
will be determined from the condition $m(T)=n$. 
In the opposite case where $m(T_{1,rsb})<n$,
that temperature is just $T_{1,rsb}$ itself.

On the other hand, in case $n<1$ the physical part is consistent 
only with $n<m<1$. Namely, for $n<1$ the possible transition point 
must be determined from the condition $m=1$.
The transition from paramagnet to RSB spin-glass with vanishing 
lower plateau with $m=1$ at the critical point has been found in 
\cite{crisanti1} for $n=0$. It occurs at the temperature
\begin{equation}
\label{xwz3}
T_{2,rsb}=J\sqrt{\frac{p}{2(1+c)} \left (\frac{1+c}{ c}\right) ^{2-p} }.
\end{equation}
At this  point $q_1$ jumps from zero to
\begin{equation}
\label{xwz4}
q_{1}(T_{2,rsb})=\frac{c}{c+1},
\end{equation}
and it goes monotonously to unity when $T$ tends to zero.
The transition  is intermediate between first-order and second-order: the
order parameter has a jump but the energy and entropy are continuous,
 as we see from Eqs. (\ref{kamo1}-\ref{kamo3}). 
 
\begin{figure}[bhb]
\vspace{0.1cm}\hspace{-1.5cm}
\vbox{\hfil\epsfig{figure=
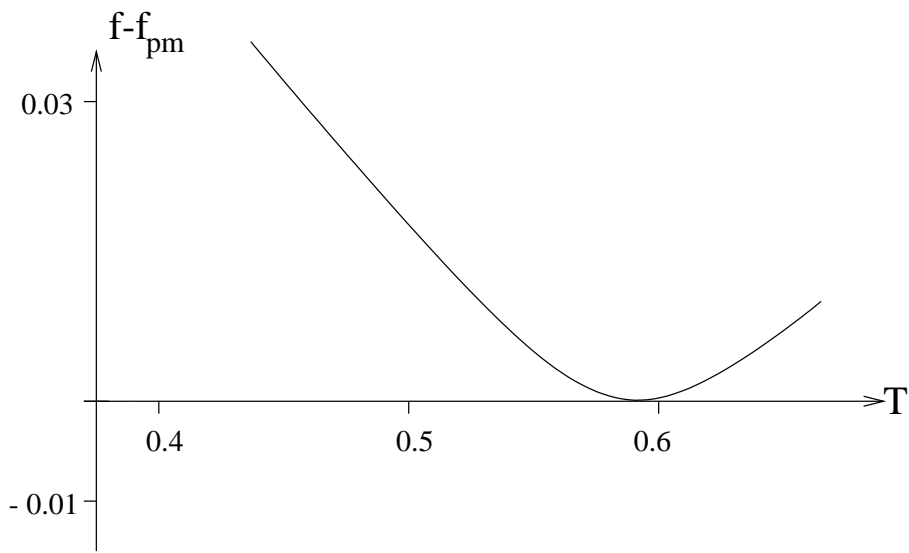,width=12cm,angle=0}\hfil}  
\vspace{0.75cm}
\caption{The free energy (subtracted the paramagnetic contribution) of the 1RSB spin-glass
phase with vanishing lower plateau  vs.
temperature ($p=3$).
For $n>1$ ($n<1$) a part of the right-hand (left-hand) 
branch of the presented curve should be chosen as the physically 
permissible one (see Eq. (\ref{prem})).
It turns out that the chosen part of the right-hand branch 
always corresponds to a metastable phase. 
}
\label{fig0}
\end{figure}

Note that the free energy in the spin-glass phase is higher than 
that of the paramagnetic state (Fig.~\ref{fig0}). 
It is usual for this type of phase
transitions \cite{parisibook}\cite{binder}\cite{pottsglass}\cite{crisanti1}. 

If $p \mapsto \infty$ then also $c \mapsto \infty$.
More concretely, in this limit it holds that
\begin{equation}
\label{xwz5}
\frac{J^2}{2T_{2,rsb}^2}\sim (\ln c-1)\exp \left (\frac{1}{\ln
c-1}\right )\mapsto \infty \end{equation}
\begin{equation}
\label{xwz6}
q_1\sim 1-\frac{1}{c}
\end{equation}
This behavior in the large-$p$ limit is in the sharp
contrast with the case of $p$-interaction  Ising spin-glass
\cite{gardner}, where
phase transition point is finite when $p\mapsto \infty$.

Let us now consider the zero temperature behavior of the solution with
fixed but not very large $p$. A simple analysis shows that in this case
\begin{equation}
\label{xwz8}
1-q_1\sim \frac{T}{J}\sqrt{\frac{2}{p(1+c)}},
\end{equation}
\begin{equation}
\label{xwz9}
m\sim \frac{T}{J}c\sqrt{\frac{2}{p(1+c)}},
\end{equation}
\begin{equation}
\label{fO0}
f(T\mapsto 0)=-\frac{J(c+p)}{\sqrt{2p(1+c)}}.
\end{equation}
Note again that the free energy of the solution remains finite 
in the zero-temperature limit in contrast to the paramagnetic 
free energy which tends to minus infinity.

The considered solution 
is stable and all relevant eigenvalues of the
Hessian are nonzero at the phase transition point.
In particular, for the relevant eigenvalue: (see \cite{crisanti1}  
for the derivation)
\begin{equation}
\label{mumu1}
\La _1=-\mu (p-1)q_1^{p-2}+\frac{1}{(1-q_1)^2},
\end{equation}
we get from Eq. (\ref{ogul3}):
\begin{equation}
\label{mumumu1}
\La _1=\frac{1}{(1-q_1)^2}\left (1-\frac{p-1}{1+c}\right )>0.
\end{equation}
The eigenvalue (\ref{mumu1}) describes fluctuations deep inside a pure state.
As usual, critical slowing down is absent at this static first order
 phase transition.
Later we shall show that in the long-time dynamics there occurs a 
marginal stability for all $n<1$.

\begin{figure}[bhb]
\vspace{0.1cm}\hspace{-1.5cm}
\vbox{\hfil\epsfig{figure=
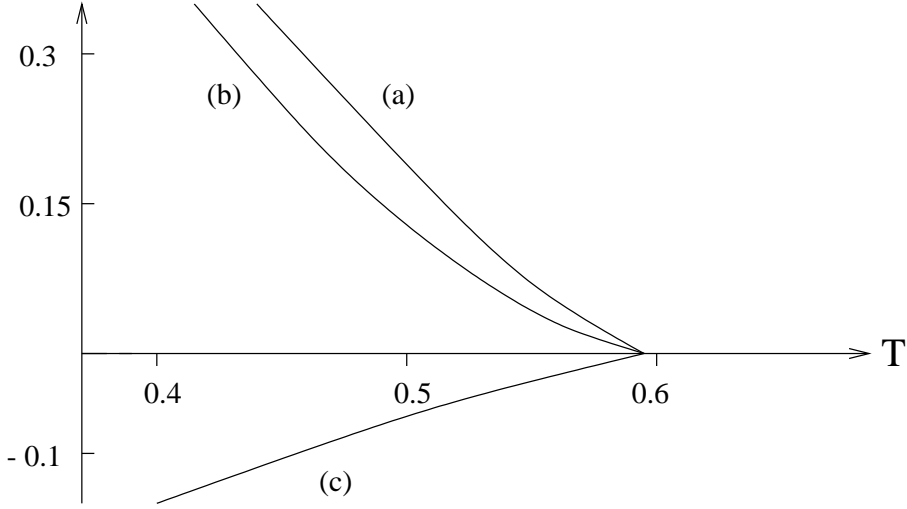,width=12cm,angle=0}\hfil}  
\vspace{0.75cm}
\caption{The entropies (subtracted the corresponding paramagnetic 
contributions; 
see Eqs. (\ref{kamo2}), (\ref{kamo3})) of the 
1RSB spin-glass
phase with vanishing lower plateau  vs.
temperature ($p=3$, $n=0.5$).
(a): $s_{\sigma}$, entropy of the spins; (b): $s$, the total entropy; (c): 
$s_{J}$, entropy of the coupling constants.
The curves  start at the transition point $T=T_{2,rsb}$. 
}
\label{fig00}
\end{figure}

It is of interest to discuss the behavior of entropies
 $s_{\sigma }$, $s_J$, and $s$ in the vicinity of the transition
 point $T_{2,rsb}$.
In particular, we have from Eqs. (\ref{kamo1}-\ref{kamo3})
\begin{equation}
\label{kamo222}
2 s_{\sigma }=\ln (1-q_1)
+\frac{1}{m}\ln (1+c)
+\frac{\mu}{p}(1-(1-m)q^p_1)
\end{equation}
\begin{equation}
\label{kamo333}
2 s_{J}=
\frac{n\mu}{p}(1-(1-m)q^p_1)
\end{equation}
In spite of the jump of $q_1$, they change continiously from the
 paramagnetic phase to the spin-glass one. We shall compare their 
behavior in spin-glass and
paramagnet at the same value of temperature.
For $T<T_{2,rsb}$ the entropy of the spins is higher in the spin-glass phase: 
$s_{\sigma }>s_{\sigma}(q_1=0)$. On the other hand, the entropy of
the coupling  constants is lower: 
$s_{J}>s_{J}(q_1=0)$, as is obvious from Eq. (\ref{kamo333}).
Consequently, the total entropy $s=s_{\sigma}+s_J$ is also higher
 in the spin-glass phase as compared
to its value for $q_1=0$. The behavior of the entropies is presented 
in Fig.~\ref{fig00}.

Further properties of the RSB spin-glass phase (in particular, for $n>1$) 
will be discussed later in the context of construction the complete
phase diagram.

\subsubsection{Replica symmetry breaking with non-vanishing lower
plateau.}
\label{nonvanish}
\  $ $ \
A solution with $q_0$ emerging smoothly from $q_0=0$ is not compatible with
the saddle-point equations (\ref{d7}).
Therefore we investigate a continuous transition from the
RS solution with $q\equiv q_1>0$ and a RSB soluiton with $q_1>0$,
$q_0>0$ but small $\De q=q_1-q_0$.
With this assumption we have  the following equations at the transition point
\begin{equation}
\label{a100}
\mu _c(p-1)q_c^{p-2}=\frac{1}{(1-q_c)^2}
\end{equation}
\begin{equation}
\label{a11}
\mu _c q_c^{p-2}=\frac{1}{(1-q_c)(1+(n-1)q_c)}
\end{equation}
The first of these expresses that the replicon eigenvalue $\Lambda_1$ of a
RS solution
with $q=q_1$ vanishes. 
The transition point is thus given by
\begin{equation}
\label{a101}
T=Jn\sqrt{ \frac{p}{2}\frac{(p-2)^{p-2}(p-1)}{(p-2+n)^p}}
\end{equation}
In the limit $p\mapsto \infty $ this transition point remains finite,
and for any $p$ it tends to zero if $n$ tends to zero.
The order parameters $q_0=q_1$ have the
following jump at the phase transition point
\begin{equation}
\label{a12}
q_c=\frac{p-2}{p-2+n}
\end{equation}
As we see, it is the same point where
stability of the RS spin-glass solution is restored.

There is a useful parametrization of Eqs. 
(\ref{d7}), (\ref{d8}), (\ref{a10}): if we introduce
\begin{equation}
\label{oo1}
x=\frac{q_0}{q_1}, \ \
c=\frac{m\De q}{1-q_1},
\end{equation}
then we have
\begin{equation}
\label{oo}
\frac{1-x^p-px^{p-1}(1-x)}{p(1-x^{p-1})(1-x)}=\frac{
(1+c)\ln (1+c)-c}{c^2}
\end{equation}

Using Eqs. (\ref{oo1}-\ref{oo}) we get at the transition point:
\begin{equation}
\label{a122}
m_c=\frac{n}{2}
\end{equation}
It can be proved that $m$ decreases with decreasing of temperature, 
and $m=0$ when $T=0$.
According to our discussion at Eqs. (\ref{prem}), (\ref{m}) it means
that a 1RSB phase with non-vanishing lower plateau cannot be considered
as a physical one, at least not for $n<2$. For $n>2$ and then $m(T) >1$ 
this solution also cannot be considered as a physical one, because
then the condition $q_1>q_0$ is violated 
(this condition does not depend on $n$, and it is 
neccessary for the interpretation of $q_0$ and $q_1$ as overlaps). 

This behavior is in the sharp contrast with SK model \cite{kondor}
 \cite{dotsenko} where the main effect of finite $n$ is to 
introduce a non-vanishing lower plateau. 
This plateau increases with $n$, and replica symmetry
 breaking disappears at some critical value.
It should be noted that in our case only the property (\ref{m}) 
forbids the existence of the considered phase for $n<2$.
All other requirements are satisfied: it is stable, and has 
well-defined free energy. Stability can be checked 
by positivity of the following
eigenvalues 
\begin{equation}
\label{mu1}
\La _1=-\mu (p-1)q_1^{p-2}+\frac{1}{(1-q_1)^2},
\end{equation}
\begin{equation}
\label{mu2}
\La _0=-\mu (p-1)q_0^{p-2}
+\left (\frac{1-q_1+nq_0}{(1-q_1)(1-q_1+m\De q+nq_0)}\right )^2
\end{equation}
We should mention the possibility of more general replica symmetry
 breaking solutions in our model. For $n=0$ it was proven 
\cite{crisanti1} that 1RSB solutions are the most general RSB
ones, and more orders of RSB are impossible. This proof can be
 generalized also for $n>0$. The physical meaning of this 
statement is that a finite $n$ introduces correlations between 
coupling constants partially removing frustrations. 
In other words, it cannot lead to more complicated phase space with more
orders of RSB.
\subsection{The static phase diagram.}
\label{phasephase}

In this subsection we construct the phase diagram of our model by
 considering all physically relevant
solutions (it means stable and with correct probability of overlaps 
(\ref{prem})): paramagnet, RS spin-glass, 
RSB spin-glass with vanishing lower plateau.

\subsubsection{The case $n<1$.}
\label{n<1}

First we discuss the case $n<1$.
For high temperatures the system is in the paramagnetic phase.
If temperature decreases, then at $T=T_{2,rsb}$ (see Eq. (\ref{xwz3})) the
RSB spin-glass with vanishing lower plateau appears. Its free energy
is greater than the paramagnetic 
one (see Fig.~\ref{fig0}) 
but, nevertheless, it is chosen as the relevant phase. 
It is the usual choice, and a possible argument is a hypothesis about a 
non-perturbative instability of the paramagnetic state below
$T_{2,rsb}$.  As far as we know, there is no
convincing proof of this statement. There is only some hint gained
from an analysis of finite-size corrections in Potts glass \cite{pottsglass}.
As we have mentioned already, this phase transition is first-order with 
respect to the order parameter but second-order with respect to
derivatives of free energy. 
In particular,  the latent heat (the 
difference between the energies of the high temperature and the low 
temperature phase at the transition point) vanishes.

\begin{figure}[bhb]
\vspace{-0.5cm}\hspace{-1.5cm}
\vbox{\hfil\epsfig{figure=
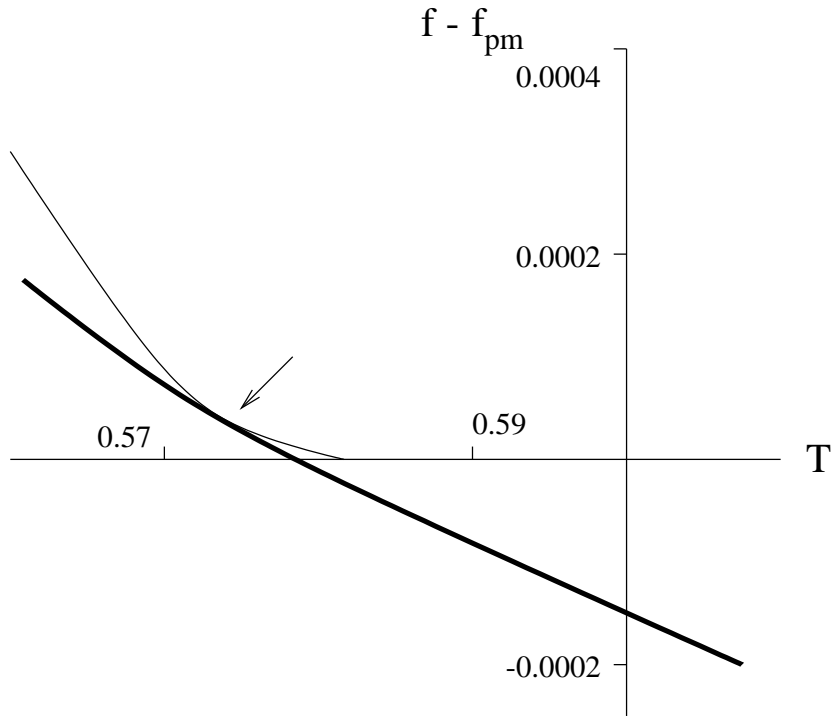,width=11cm,angle=0}\hfil}  
\vspace{0.75cm}
\caption{Free energies (subtracted the paramagnetic free energy) vs.
temperature ($n=0.955$, $p=3$).
Thick line: the RS spin-glass phase; normal line: the RSB spin-glass
phase with vanishing lower plateau. The arrow denotes the 
phase transition point occurring at 
$T=T_{3,rs}$ from the 1RSB spin-glass phase to the RS one. 
}
\label{fig1}
\end{figure}

\begin{figure}[bhb]
\vspace{0.1cm}\hspace{-1.5cm}
\vbox{\hfil\epsfig{figure=
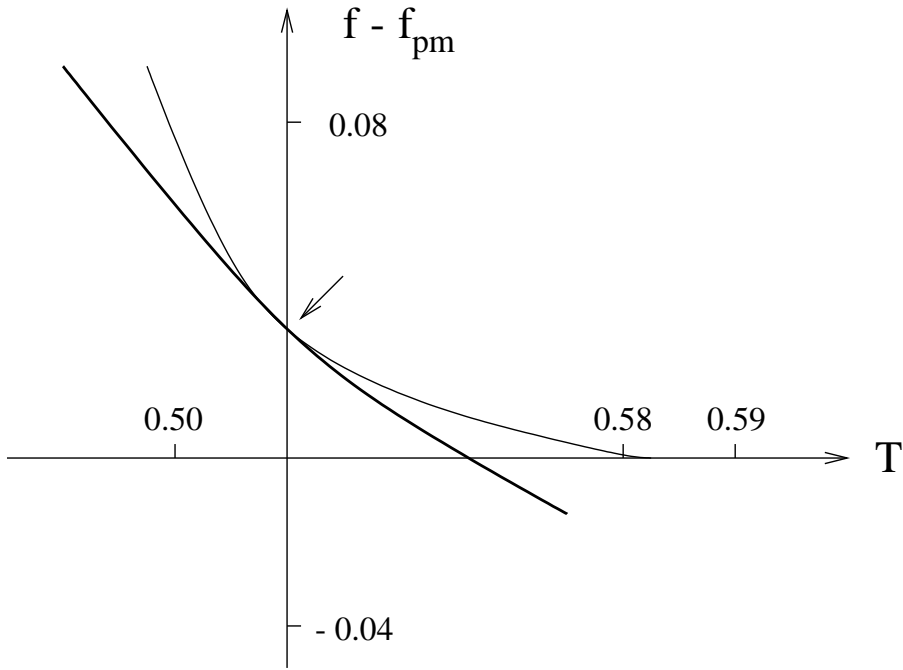,width=12cm,angle=0}\hfil}  
\vspace{0.75cm}
\caption{Free energies (subtracted the paramagnetic contribution) vs.
temperature ($n=0.8$, $p=3$).
Thick line: the RS spin-glass phase; normal line: the RSB spin-glass
phase with vanishing lower plateau. The arrow denotes the phase transition point at $T=T_{3,rs}$.
}
\label{fig2}
\end{figure}

At the present stage we shall go back to the RS spin-glass 
solution, and analyze its free energy. Recall that this phase is stable for 
$T<T_{rs,st}$ (see Eq. (\ref{rsstab1})), and for some range of $n$ 
we have $T_{rs,st}>T_{2,rsb}$ (see Fig.~\ref{fig1}).
In particular, the sign of
\begin{equation}
\label{free-1} \De f_{rs}\equiv  
f_{rs}-f_{para}
= -\frac{T}{2}\ln (1-q)-\frac{T}{2n}\ln \left (1+\frac{nq}{1-q} \right )
+\frac{\beta J^2}{4}(1-n)q^p
\end{equation}
should be checked.
In this respect the interval $0<n<1$ is divided into two subintervals.
For $n_0<n<1$ (where $n_0$ is some positive value, to be discussed later) 
it holds that $\De f_{rs}(T_{rs,st})<0$.
Taking into account that at $T=T_{rs,st}$ RS spin-glass first appears
as a physical solution, we conclude that in this range of temperatures 
it should be considered as metastable with respect to the paramagnet,
in spite of its lower free energy. The opposite point of view will 
mean that $T=T_{rs,st}$ must be considered as
the point of a phase transition which is meaningless, because $
\De f_{rs}(T_{rs,st})\not =0$. If temperature is decreasing further 
we get the point $T=T_{2,rs}$ where
$\De f_{rs}(T_{2,rs})=0$, and $\De f_{rs}(T_{rs,st})>0$ for 
$T_{2,rs}>T$. This behavior of the free energy is presented 
by Figs.~\ref{fig1}, \ref{fig2}. However, 
this point is always lower than the transition temperature from 
the paramagnet into the RSB spin-glass phase with vanishing lower plateau: 
$T_{2,rsb}>T_{2,rs}$. (On the other hand
$T_{2,rsb}\leq T_{2,rs}$ for $1\leq n$, but this case is physically different
and will be discussed a bit later.) Therefore, the temperature 
$T=T_{2,rs}$ is not considered as a true
phase transition point towards the most stable phase. 
Certainly, it is only the point where RS spin-glass phase becomes more 
stable than the paramagnetic state. 

For $n<n_0$ one has $\De f_{rs}>0$, so the RS spin-glass state 
at the beginning appears
with higher free energy than the paramagnet.  
$n_0$ is defined by the condition $\De f_{rs}(T_{rs,st})=0$.
In Table~I
we represent the values of $n_0$ for different $p$ . In particular, we 
see that $n_0(p)$ decreases with increasing $p$. 

\vbox{
\begin{table}[bhb]
{\footnotesize
$$
\begin{array}{||c|c||}
 \hline n_0 & p \\  \hline
0.68632        & 3           \\  \hline
0.61605        & 4           \\  \hline
0.53108        & 6           \\  \hline
0.36803        &18           \\   \hline
0.31670       &30            \\   \hline 
\end{array}
$$
}
\caption{ 
$n_0$ for different $p$ ($J=1$).
}
\end{table}
}

According to Eqs. (\ref{prem}), (\ref{m}) for $n<1$ any 1RSB spin
glass cannot exist as a physical one if $m<n$. From subsection \ref{vanish}
we know that for the RSB spin-glass with vanishing lower plateau 
$m$ monotonically decreases with temperature from $m=1$ at $T=T_{2,rsb}$ to
$m=0$ at $T=0$. It means that a phase transition should exist from 
a RSB  spin-glass with vanishing lower plateau to a RS spin-glass
at a temperature $T_{3,rs}$ defined by
\begin{equation}
\label{ja}
m(T_{3,rs})=n. 
\end{equation}

\vbox{
\begin{table}[bhb]
{\footnotesize
$$
\begin{array}{||c|c||}
\hline n & T_{3,rs}    \\ \hline
0.90       & 0.5575            \\ \hline
0.80       & 0.5240           \\ \hline
0.70       & 0.4805           \\ \hline
0.60       & 0.4413           \\ \hline
0.25       & 0.2326             \\ \hline
\end{array}
$$
}
\caption{ The transition temperature
$T_{3,rs}$ from the 1RSB spin-glass
phase to the RS one, for different $n$ ($n<1$), and $p=3$
($J=1$). }
\end{table}
}

As Eqs. (\ref{dopdop1})-(\ref{a10}) show, the order 
parameters are equal  at this point: $q_1=q$.
It is also easy to show that the free 
energies, energies and entropies 
$s_{\sigma}$, $s_J$ of the corresponding phases are also equal.
Thus, at $T=T_{3,rs}$ the order parameter $q_0$ jumps from zero to 
$q_1$, ensuring the replica-symmetric 
behavior for $T<T_{3,rs}$. On the other hand, $q_1$ changes continuously.
Both the RS spin-glass and the RSB spin-glass with vanishing lower 
plateau are stable at and around $T=T_{3,rs}$.
In this sense the transition at $T=T_{3,rs}$ is very similar to the 
transition occurring from the paramagnetic state at $T=T_{2,rsb}$, 
where $q_0$ changes continuously
(i.e., remains zero) but $q_1$ has a jump.
It is worth to note
that for $T<T_{3,rs}$ the free energy of the
RS spin-glass is lower than the free
 energy (more exactly its analytical continuation) of 
the RSB spin-glass with vanishing lower plateau.
Different values of $T_{3,rs}$ are presented in Table~II.

Summarizing the presented facts and arguments, we conclude that 
when temperature is decreasing for fixed $n<1$
there are only two temperatures of true phase transitions: 
1) $T=T_{2,rsb}$ is the transition point
from the paramagnet to the RSB spin-glass with vanishing lower plateau. 
2) $T=T_{3,rs}$ is the transition point
from the RSB spin-glass into the RS one. 
These transitions are quite close by their physical meaning. 
The main distinction is in the difference between free energies
of the corresponding low-temperature phase and the high-temperature one 
(or rather its analytical continuation to lower temperatures).
The temperatures $T_{rs,st}$ and $T_{2,rs}$ 
also have certain physical meanings, but 
are not considered as true transition temperatures.

\begin{figure}[bhb]
\vspace{0.15cm}\hspace{-1.5cm}
\vbox{\hfil\epsfig{figure=
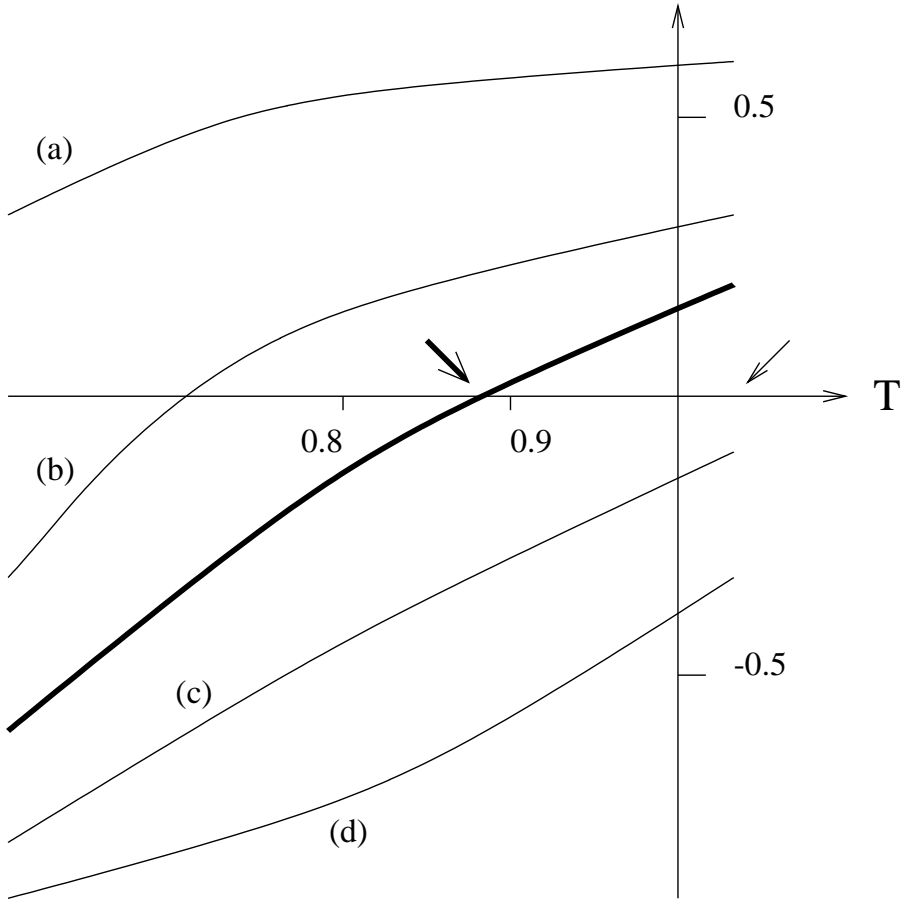,width=12cm,angle=0}\hfil}  
\vspace{0.75cm}
\caption{Thermodynamical functions 
(minus the corresponding paramagnetic contributions) 
of the RS spin-glass ($n=4$, $p=3$) vs.
temperature. The thick arrow denotes the phase 
transition point from paramagnet to the RS spin-glass phase
at $T=T_{2,rs}$.
The normal arrow denotes the temperature $T=T_{1,rs}$, 
where the RS spin-glass solution appears first as a metastable phase.
Thick line: free energy; (a): $s_J$, entropy of the coupling constants; 
(b): $s$, the total entropy; (c): $u_{rs}$,
mean energy; (d): $s_{\sigma}$, entropy of the spins.
}
\label{fig3}
\end{figure}

\subsubsection{The case $n>1$.}
\label{n>1}

In the case $n>1$ there are three relevant solutions:
the RS spin-glass ($q>0$), the paramagnet ($q=0$), and 
the RSB spin-glass with vanishing lower plateau in the region $1<m<n$.
Recall that this last state formally exist for $T<T_{1,rsb}$ 
(see Eq. (\ref{ogul44})), but the region of its actual existence
must be chosen according to the value of $n$ 
as it has been done for $n<1$. Doing so we 
immediately conclude that for $n>1$ the RSB spin-glass with vanishing 
lower plateau cannot be considered as the most stable one, 
because its free energy is always higher than
the free energy of the paramagnetic state (Fig.~\ref{fig0}). 
It can be viewed only
as a metastable one in the above-mentioned region. 
The behavior of the corresponding free energy is presented in
Fig.~\ref{fig0}.
Therefore, our attention will be restricted only to the paramagnet
and the RS spin-glass phase.

According to our discussion in subsection \ref{subsec:rs}, a RS
spin-glass phase first appears at $T=T_{1,rs}$ (Eq. (\ref{serso1}))
 but its free energy at this point is higher than the paramagnetic 
one. These free energies are 
equal when $T=T_{2,rs}$, and for $T<T_{2,rs}$ the RS spin-glass phase
 has lower free energy.
Thus, the temperature $T_{2,rs}$ is the true thermodynamical phase 
transition point. This transition is first order, and it is connected
 with a jump of the 
order parameter $q$. It is obvious from Eq. (\ref{kamo11}) that for 
$T<T_{2,rs}$ the mean energy of the spin-glass phase is also lower than 
the paramagnetic one.
It means that the latent heat is positive as for (usual) first-order phase
 transitions in equilibrium systems. Values of $T_{1,rs}$, and
$T_{2,rs}$  are represented in Table~III.
In particular, this scenario of phase transition is the usual one for
multi-spin interaction ferromagnets \cite{nagaev}\cite{fisher}, 
for some metamagnetic materials \cite{nagaev} or for phase transitions
in a compressible lattice \cite{pokrov}.
For such a system the jump of entropy at the transition point
is negative (i.e., a low-temperature phase has lower entropy) according
to the usual relation $F=E-TS$ between free energy, energy and entropy.
At this stage it should be recalled again that our system is not in
the usual equilibrium, and we are considering phase transitions in the
nonequilibrium steady state.
In particular, we have the basic relation (\ref{terra1}) between 
free energy, energy, and entropies. At the transition point 
$T=T_{2,rs}$ this relation can be written as
\begin{equation}
\label{ja1}
\Delta u =T\Delta s_{\sigma }+T_J\Delta s_J, 
\end{equation}
where $\Delta u =u _{rs}-u _{pm}$, $\Delta s_{\sigma }=
 s_{\sigma , rs} - s_{\sigma , pm}$, and
$\Delta s_J = s_{J, rs} - s_{J , pm}$ are differences between 
the corresponding quantities 
of the RS spin-glass and the paramagnet, and $u _{rs}$, $s_{\sigma , rs}$,
$s_{J , rs}$ are defined by Eqs. (\ref{kamo11})-(\ref{kamo33}). 
Further, $T$ and $T_J$ are connected through $T=T_{2,rs}$. 
In particular, positive latent heat means
$\Delta u<0$, and consequently $T\Delta s_{\sigma }+T_J\Delta s_J<0$. 
However, it is interesting to know signs of $\Delta s_{\sigma }$, 
$\Delta s_J$, and $\Delta s=\Delta s_{\sigma }+\Delta s_J$ separately 
because these quantities have independent physical meanings,
extensively  discussed in section \ref{sec:2T}. We get
\begin{equation}
\label{ja2}
\Delta s>0, \, \Delta s_J>0, \, \Delta s_{\sigma }<0. 
\end{equation}
The behavior of $\Delta s$ is in the sharp contrast with the usual,
equilibrium first-order transitions. This somewhat surprising fact 
should deserve  further attention.
The behavior of various thermodynamical quantities near $T_{2,rs}$ is 
presented in Fig.~\ref{fig3}.
The difference between spins entropies $\Delta s_{\sigma }$ becomes 
positive starting from some temperature lower than $T_{2,rs}$. 
In other words, for sufficiently
low temperatures both entropies are higher in the spin-glass phase. 
Let us recall in this context that free energy and mean energy of the
 spin-glass phase are lower than 
the paramagnetic 
ones for sufficiently low temperatures.

Positivity of the latent heat, and the result (\ref{ja2}) holds also for
 the first-order phase 
transitions occurring for $p=2$, $n>2$ (see section \ref{p=2}). 
In the case of positive but small $n-2$ we can get analytical 
expressions for $\Delta s_{\sigma }$, $\Delta s_J$, and $\Delta s$ 
using Eqs. (\ref{puk1}-\ref{puk3}). The jump $q_{2,rs}$ of the order
 parameter is positive but small at $T=T_{2,rs}$,
and we get to leading order
\begin{eqnarray}
\label{puk4}
&&\Delta s_{\sigma }=-\frac{1}{2}q^2_{2,rs}+{\cal O}(q^3_{2,rs})<0, \  
\ \Delta s_J=\frac{1}{2}q^2_{2,rs}+{\cal O}(q^3_{2,rs})>0, \nonumber \\
&&\Delta s=\frac{5}{24}q^3_{2,rs}+{\cal O}(q^4_{2,rs})>0
\end{eqnarray}
This is an analytical illustration of the more general result (\ref{ja2}).

The special attention should be devoted to the case $n=1$. 
The free energy, and the energy of the RS spin-glass coincide with
 their analogs for the paramagnetic phase. However,
from the point of view of the order parameter, the phase 
transition occurs to the RS spin-glass phase 
(when decreasing temperature for a fixed $n=1$). 
Further, $q$  increases monotonically from $q=(p-2)/(p-2+n)$ at 
$T=T_{2,rsb}=T_{2,rs}=T_{3,rs}$ to $q=1$ at $T=0$. 
The phase transition is first-order for $p>2$, and 
second-order for $p=2$.
It is interesting that $s_{\sigma }$ and $s_{J}$ depend 
on $q$ even for $n=1$, however, they compensate each other so that 
their sum $s=0$. 
The final phase diagram is presented in Fig.~\ref{phase}.

\begin{figure}[bhb]
\vspace{0.15cm}\hspace{-1.5cm}
\vbox{\hfil\epsfig{figure=
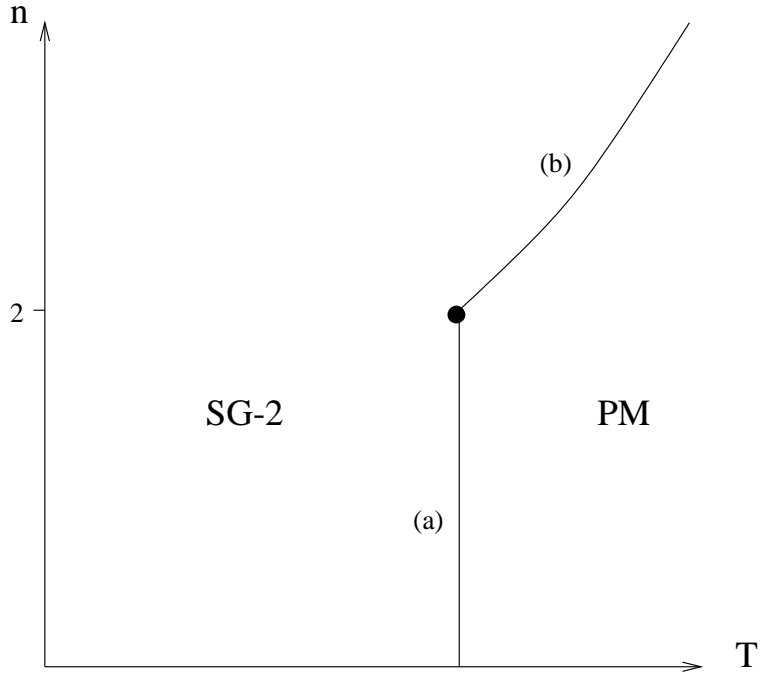,width=10cm,angle=0}\hfil}  
\vspace{0.75cm}
\caption{The static phase diagram of the model 
in the $n$ | $T$ plane for $p=2$.
The paramagnetic phase is denoted by PM,
and SG-2 means 
the RS spin-glass phase.  (a)
$T=J$, the line of the second-order transitions; 
(b) $T=T_{2,rs}(n)$, the line of the 
first-order transitions with a positive jump of the total entropy. 
The thick dot indicates the multicritical point.
}
\label{phasep2}
\end{figure}
 
\begin{figure}
\vspace{0.15cm}\hspace{-1.5cm}
\vbox{\hfil\epsfig{figure=
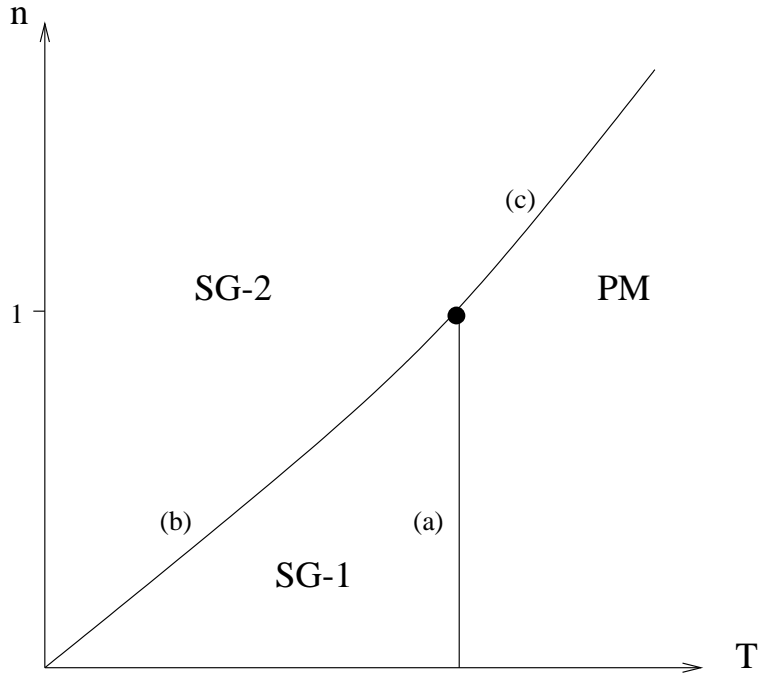,width=10cm,angle=0}\hfil}  
\vspace{0.75cm}
\caption{The static phase diagram of the model in the $n$ | $T$ plane for $p>2$.
The paramagnetic phase is denoted by PM,
and SG-1, SG-2 mean
correspondingly the RSB spin-glass phase with vanishing lower plateau, and
the RS spin-glass phase. (a)
$T=T_{2,rsb}(p)$; (b) $T=T_{3,rs}(n,p)$; (c) $T=T_{2,rs}(n,p)$. 
The lines (a) and (b)
indicate first-order transitions without latent heat; (c) indicates the first-order transition 
with the positive latent heat, and a positive jump of the total entropy.
The thick dot indicates the multicritical point.
}
\label{phase}
\end{figure}

\section{Ergodic dynamics} 
\label{sec:dyn}

\  $  $ \
At the present section we consider the ergodic (or time-translation
invariant) dynamics 
of our two-temperature model.
As can be expected from our experience acquired from the 
study of statics, if $n<1$ 
ergodicity holds at relatively short but infinite times where
RSB is not relevant 
(of course, that restriction of times is not essential if 
temperatures is high enough). 
On the other hand, if $n>1$ the assumptions connected with
ergodicity are correct for any
temperature. 
Indeed, in the last case the RS assumption adequately describes the
system's state for all temperatures.
Thus, as opposed to the case of usual mean-field spin-glass systems
(with $n=0$), in our case ergodicity can have a wider range of applicability.
In particular, in the present section we compare the predictions 
of the long-time dynamics with 
the static ones, 
as well as investigate effects
connected with very large observation times.

Eqs. (\ref{5}, \ref{6}) have the following form in the time-translation
invariant regime
\begin{eqnarray}
\label{risk5}
& &(\pdt +r)C(t)=\frac{pJ^2}{2\Gamma _J} \int_{0}^{\infty}
d\tdpr e^{-\tdpr /\tau _J} C^{p-1}(\tdpr )C(t-\tdpr )\nonumber \\
&&
+\frac{pT_J J^2}{2v} (p-1) \int_{0}^{\infty} d\tdpr e^{- \tdpr /\tau _J} 
C^{p-2}(
\tdpr)G(\tdpr ) C(t-\tdpr )\nonumber \\
&&
+\frac{pT_J J^2}{2v}\int_{0}^{\infty } d\tdpr  e^{- (t+\tdpr )/\tau _J}
C^{p-1}(t+\tdpr
)G(\tdpr ),
\end{eqnarray}
\begin{eqnarray}
\label{risk6}
&  &(\pdt +r)G(t)=\frac{pJ^2}{2\Gamma _J} \int_{0}^{t}
d\tdpr e^{- (t-\tdpr )/\tau _J} C^{p-1}(t-\tdpr )G(\tdpr )\nonumber \\
& &+\frac{pT_J J^2}{2v} (p-1) \int_{0}^{t} d\tdpr e^{-(t-\tdpr )/\tau _J} 
C^{p-2}(t-\tdpr)G(t-\tdpr ) G(\tdpr ).
 \end{eqnarray}

\subsection{Adiabatic dynamics: not very large observation times.}
We come to the adiabatic statics, which has been considered in section \ref{sec:2T},  
from the dynamics
if $\tilt \mapsto \infty $ before $t\mapsto \infty $, so that
the usual form of  FDT
\begin{equation}
\label{fdtold}
\pdt C(t)= T(-G(t)+G(-t))
\end{equation}
can be used.
Indeed, in this case for all $t\ll \tau _J $ the coupling constants are fixed, and the violation
of the detailed balance condition (which arises in account of a difference between temperatures) 
cannot be effective.
Thus we have the equation
\begin{eqnarray}
\label{b5}
&  &(\pdt +r)C(t)=\frac{pJ^2}{2\Gamma _J}
\int_{0}^{\infty }
d\tdpr e^{-\tdpr /\tau _J }
C^{p-1}(\tdpr )C(t-\tdpr )\nonumber \\
&  &+\frac{pT_J J^2}{2v} (p-1)
\int_{0}^{\infty }
d\tdpr e^{- \tdpr/\tau _J }
C^{p-2}(\tdpr)C(t-\tdpr )G(\tdpr )
\nonumber \\
&  &+\frac{pT_J J^2}{2v} \int_{0}^{\infty }
d\tdpr e^{- \tdpr /\tau _J } G(\tdpr )C^{p-1}(t+\tdpr )
\end{eqnarray}
In the last two integrals we can take $e^{-\tdpr /\tau _J }\sim 1$ by considering
the relevant domain of the integration.
In the first integral we find  by a changing variables
\begin{equation}
\label{bb1}
\frac{1}{\Gamma _J}
\int_{0}^{\infty }
d\tdpr e^{- \tdpr /\tau _J }
C^{p-1}(\tdpr )C(t-\tdpr )\mapsto \frac{1}{v} q^p,
\end{equation}
where
\begin{equation}
\label{rusalka1}
q={\rm lim}_{t\to \infty} ({\rm lim}_{\tau _J\to\infty} C(t))
\end{equation}
Now for investigating phase transitions we take our usual restriction to the
parameters of the model:
$v= T_J$, $n\equiv T/T_J$ is fixed, and $T$ is
varying. Thus $\mu =p\beta ^2J^2/2$.
The static limit of (\ref{b5}) gives us
the RS equation (\ref{a4}).
Indeed, for determination of $q$ we can take the  two limits:
$t=0$ where $C(0)=1$, $\pdt C(t)|_{t=0}=-T$, and $t\mapsto \infty$ where $C=q$.
For these cases we have the following equations
\begin{equation}
\label{heihei1}
\beta r-1=\mu n q^p+\mu (1-q^p)
\end{equation}
\begin{equation}
\label{heihei2}
\beta rq=\mu n q^p +\mu (q-q^p)+\mu q^{p-1}(1-q)
\end{equation}
These equations reproduce eq. (\ref{a4}), but
now as a result of long-time dynamics. As was shown in 
section
\ref{sec:stat}, the non-zero solution
of this equation correctly describes the low-temperature phase of the system
for $n>1$. For obtaining the RSB equations from
the dynamics a non-ergodic regime should be considered;
it will be considered elsewhere. 

We see that the replica-symmetric equation for the order parameter
can be found from the ergodic dynamics.
But in  systems with first order phase transition predictions
for long-time dynamics and statics can be different.
 What is the meaning a dynamic phase transition? 
Our starting point involved
Langevin equations which are first order with respect of time
(more exactly, they are the overdamped limit of the
true Langevin equations). Thus our dynamics is purely relaxational.
In other words, we must have: $\pdt C(t)\leq 0$.
Of course,  if the temperature is high enough then $\pdt C(t) <0$.
We can determine the dynamic phase transition point
by the following two conditions: it is the temperature $T_d$
where $\pdt C(t)=0$, and for all $T<T_d$ we then have $\pdt C(t)>0$.
If such temperature exists, then a contradiction arises because
we know that the dynamics is purely relaxation.
The only way to solve this contradiction is to assume that the ergodicity 
is broken
for all $T<T_d$.

By the same methods as in \cite{crisanti2} we have from (\ref{b5})
\begin{eqnarray}
\label{qurd1}
&  & (\pdt +r)C(t)=\mu T  C(t)+(\beta r-\mu -1)T+\mu T 
C^{p-1}(t)(1-C(t)) \nonumber \\
&  &-\mu T \int_{0}^{t } d\tdpr
\pa _{\tdpr }C(\tdpr)[C^{p-1}(t-\tdpr)-C^{p-1}(t )]
\nonumber \\
&  &+\frac{pJ^2}{2\Gamma _J} \int_{0}^{\infty }
d\tdpr e^{- \tdpr /\tau _J } C^{p-1}(\tdpr )(C(t-\tdpr )-C(\tdpr ))
\end{eqnarray}
We find that for large $t$ and $n<1$ 
the dynamical stability condition
$\pa _{t}C(t)\leq 0$ is equivalent to the following two equations,
which were obtained  already in \cite{crisanti2},
\begin{eqnarray}
\label{d22}
&&
\beta r-\mu =\bar{r}(C(t)) \nonumber \\
&&
\bar{r}^{\pr }(C(t))=0,
\end{eqnarray}
where 
\begin{equation}
\label{d21}
\bar{r}(C(t))= \frac{1}{1-C(t)}-\mu C^{p-1}(t)
\end{equation}
These equations predict a phase transition at the temperature
(assuming that it is the highest temperature where a phase transition
from paramagnet occurs)
\begin{equation}
\label{ali1}
T_d^2=\frac{pJ^2}{2}\frac{(p-2)^{p-2}}{(p-1)^{p-1}}
\end{equation}
with jump
\begin{equation}
\label{ali2}
q_d=\frac{p-2}{p-1}
\end{equation}
The standard interpretation of this phase transition is the following
\cite{crisanti2,CSTAP,theoprl1,cug}.
The dynamics is investigated in the limits: the initial time $\to -\infty$
after the thermodynamic limit $N\to \infty$.
In contrast, in the purely static investigation by means of the Gibbs
distribution, the opposite order of limits is implicated.
Thus in the long-time dynamics the system can be blocked in some regions of
the phase space.
It was shown that, indeed, the dynamical phase transition is induced by the
highest metastable TAP states \cite{CSTAP} which are separated by each 
other by means of infinite
barriers (in the thermodynamic limit).
The dynamical phase transition
can be reflected by  an other approach also \cite{theoprl1}:
 As we have seen in section \ref{sec:stat}, the relevant
eigenvalue of Hessian is (\ref{a5})
$$
\La _1=-\mu  (p-1)q^{p-2}+\frac{1}{(1-q)^2}
$$
(which does not dependent on $n$). By now considering a "long-time" but not
purely static regime, the RSB parameter $m$ can be fixed by the 
condition $\La _1=0$.
In this approach (marginal replica theory) the phase transition point
coincides with the dynamical one, but the thermodynamics is different 
compared to the one which is obtained
from the long-time limit of the dynamical equations.

We have seen that the dynamical transition point is independent of $n$ 
if this temperature is larger than all other transition points
predicted by statics. A simple comparison gives the following.
For all $0\leq n< 1$ there is a difference between statics and dynamics:
$T_d>T_{2,rsb}>T_{3,rs}$.
It coincides with the fact that for this case the
relevant eigenvalue of Hessian (\ref{a5}) does not depend on $n$.
There is no such difference for $n>1$, because in this case the statics
predicts a phase transition at a greater temperature
than the dynamics: $T_{1,rs}>T_{2,rs}>T_d $. 
Indeed, in this case the replicon or ergodon 
eigenvalue $\La _1$ is not a relevant one.

We see that the existence and probably the  structure of the highest
metastable TAP states strongly depends on $n$.

\subsection{Adiabatic dynamics: very large observation times.}

In this subsection we have concentrated exclusively on
the case $n>1$; it was mentioned above that
in this case ergodicity is valid without any constraint on the
temperature or the time of observation. 
As we have seen in previous subsection, there is a spin-glass phase
if the observation time is not very large, i.e.,
if $t\ll \tau _J$. Further, the same result as predicted by the 
long-time dynamics can be obtained from the statics. 
However, for $t\sim \tau _J$ (i.e., when an observer
waits long enough) the coupling constants will begin to fluctuate 
and relax toward their steady state.
It is expected intuitively that at such observation times
the spin-glass phase will disappear, and the resulting stationary 
state will be a paramagnet. As a matter of fact, this subsection is devoted 
to confirm that point of view. 
Furthermore, we will see that there are some peculiar 
features of this process, which probably have universal character 
related to other  glassy systems.
In the concrete calculations we use the methods of \cite{crisanti2}.

In the domain $t\sim \tau _J\to \infty$ the main t
ime-scale is $\tau _J$; thus, it is natural to take the 
following form for the correlation and response functions
\begin{equation}
\label{udav1}
C(t)={\cal C}\left (\frac{t}{\tau _J}\right ), \, 
G(t)=\frac{1}{\tau _J}{\cal G}\left(\frac{t}{\tau _J}\right )
\end{equation}
(the extra factor $1/\tau _J$ for ${\cal G}$ appears on account of 
the correct dimension of that 
quantity; furthermore its necessity will be evident further). 
Let us define also an auxiliary large
time-scale $t_e$ at which the correlation function stabilizes: 
$C(t\sim t_e)=q$ (where $q$ is defined
by Eq. (\ref{a4})). Let us stress that when in the previous subsection we 
have spoken about very large $t$ where $C(t)=q$,
we meant | in the light of the present discussion | the case $t\sim t_e$.

Now our very large time-scales can be presented as: $\tau _J\gg t_e$;
 the cases $t\sim t_e$ and $t<t_e$ have been studied in the previous
 subsection, and now we are going
to consider the cases $t\sim \tau _J$, $t\gg \tau _J$. 
The corresponding equations for ${\cal C}$,
${\cal G}$ must be constructed from (\ref{risk5}, \ref{risk6}) 
in accordance of a simple physical 
picture occurring from the large separation of the local relaxation 
times. Namely, if the temporal
argument of $C(\tdpr)$, $G(\tdpr)$ is less or equal $t_e$, then the 
case of not very large 
observation times holds with all its consequences. For instance, 
the usual FDT can be used. At the
same time the kernel $\exp(-\tdpr /\tau _J)$ can be put equal to
unity, as  we have done more than once in the 
previous subsection. If the corresponding temporal argument has at
 least the same order as $t_e$ we 
should take into account that $C(t\sim t_e)=q$, and if it has the 
same order as $\tau _J$ we use 
(\ref{udav1}) accompanying with an evident consistency condition 
${\cal C}(0)=q$. 

Our transformations come to dividing 
the domains of integration in Eq. (\ref{risk5})
into two parts: from $0$ to $t_e$, and from $t_e$ to $\infty$ 
(the corresponding integrals in Eq. 
(\ref{risk6}) are divided into three parts because more accuracy is needed: 
from $0$ to $t_e$, from $t_e$ to $t-t_e$, and from $t-t_e$ to
$\infty$). 
Now every part is treated as described above. 
We omit all factors which have relatively
small  order when $\tau _J\to \infty$.
In particular, for Eq. (\ref{risk5}) ((\ref{risk6})) these factors
are of order ${\cal O}$ $(1/\tau _J)$ (${\cal O}$ $(1/\tau _J^2)$).

Finally, we get the following equations (where $s=t/\tau _J$)
\begin{eqnarray}
\label{udav5}
& &r{\cal C}(s)=\frac{pJ^2}{2v} \int_{0}^{\infty}
d\sdpr e^{- \sdpr} {\cal C}^{p-1}( \sdpr ){\cal C}(s- \sdpr)
+\frac{p(p-1)T_J J^2}{2v} \int_{0}^{\infty} d\sdpr e^{- \sdpr} 
{\cal C}^{p-2}(
\sdpr){\cal G}(\sdpr ){\cal C}(s-\sdpr )\nonumber \\
&&
+\frac{pT_J J^2}{2v}\int_{0}^{\infty } d\sdpr  e^{- (s+\sdpr )}
{\cal C}^{p-1}(s+\sdpr
){\cal G}(\sdpr ) +\frac{pT_J J^2}{2v}\beta e^{-s}{\cal C}^
{p-1}(s)(1-q) \nonumber \\
&& +\frac{pT_JJ^2}{2v}\beta {\cal C}(s)(1-q^{p-1}),
\end{eqnarray}
\begin{eqnarray}
\label{udav6}
& &r{\cal G}(s)=\frac{pJ^2}{2v} \int_{0}^{s}
d\sdpr e^{- \sdpr}{\cal C}^{p-1}(\sdpr ){\cal G}(s-\sdpr )
+\frac{p(p-1)T_J J^2}{2v} \int_{0}^{s} d\sdpr e^{-\sdpr} {\cal C}^{p-2}(\sdpr)
{\cal G}(s-\sdpr ) {\cal G}(\sdpr )\nonumber \\
&&+\frac{pT_J J^2}{2v}\beta {\cal G}(s)(1-q^{p-1})
+\frac{p(p-1)T_JJ^2}{2v}\beta e^{-s}{\cal C}^{p-2}(s)
{\cal G}(s)(1-q)\nonumber \\
&& +\frac{pJ^2}{2v}\beta e^{-s}{\cal C}^{p-1}(s)(1-q).
\end{eqnarray}
Now it is important to realize that the usual formulation of FDT
does not hold in this case.
Indeed, if the coupling constants fluctuate, then the heat current between 
the thermal baths cannot be neglected. In other words, the detailed 
balance condition is violated because there is a stationary current 
between the two heat baths, which changes its sign under time-reversal. 
Thus, we have a steady but non-Gibbsian state \cite{noneqrev}. 
Generally speaking, in such a state there does not exist any simple or even 
closed general relation between the correlation 
and response functions \cite{strat}. 
However, such a relation is possible in some particular
cases. It is interesting that in our case some generalized FDT exists,
which is, however, non-universal in contrast to the usual one. 
Furthermore, we shall see that this theorem does not depend on
secondary details of the model. As such it belongs to the 
thermodynamical  picture of the glassy state, together with the
law for the change of heat (\ref{dQ=}) and non-equilibrium
fluctuation relations~\cite{NEhren}\cite{theotwotem}\cite{Nlongthermo}.
Let us consider the generalized FDT in the form
\begin{equation}
\label{fdtnova-0}
\pds {\cal C}(s)=\tilde T(-{\cal G}(s)+{\cal G}(-s))
\end{equation}
with some unknown coefficient
$\tilde T$. After some calculations using (\ref{udav5}), (\ref{udav6})
we get 
\begin{equation}
\label{fdtnova}
\tilde T=T_J
\end{equation}
It is customary to call this the modified FDT the
``Fluctuation-Dissipation Relation'' (FDR) or just the ``modified FDT''.
In the our two-temperature dynamics without detailed balance
this is a very simple relation between the  correlation function 
and the susceptibility. It is important that the form of the
FDR does not depend on concrete characteristics of the model.

We have obtained that the coefficient of FDR (it is $T_J$ in the case 
 of Eq. (\ref{fdtnova}), and
$T$ in the case of (\ref{fdtold})) depends crucially on the
 observation time: There is a direct
correspondence between the time-scales which are considered and the
 proper temperature. Such
phenomena were predicted recently for systems with ``slow" dynamics 
\cite{Nlongthermo,cug,peliti}.
Particularly, such considerations are a possible fundament 
for generalizing the notion of temperature.

By the usual FDT (\ref{fdtold}) only an ergodic or short-time dynamics can be
monitored \cite{parisibook,binder,crisanti2}.
To take into account effects of RSB some regularization procedure is
necessary. As was shown in \cite{crisanti2} for
the $n=0$ $p$-spin spherical model the non-ergodic dynamics can be achieved by
some other, ``long-time" FDT
$$
\pdt C(t)=\frac{T}{m}(-G(t)+G(-t)),
$$
where $m$ is the RSB parameter. In the section \ref{sec:stat} we discussed the
analogy between the finite-$n$ RS
free energy and the RSB free energy with $q_0=0$. In the spirit of
 this analogy
this long-time FDT corresponds to (\ref{fdtnova})
because via the substitution $m\to n$ we go to this equation.
This analogy is not an accident, of course. It displays a deep connection
between systems with RSB and ones where different time-scales of
relaxation
 and different 
temperatures are assumed initially. It seems to us that other 
interesting results can be obtained in this way in future.

With help of the FDR we obtain the following formula for ${\cal G}(0)$:
\begin{equation}
\label{udav77}
{\cal G}(0)=\frac{\beta q(1-q)}{q-(p-2)\beta T_J(1-q)}
\end{equation}
Now the positivity of this quantity requires our old condition (\ref{rsstab}).
As we have seen, this condition is the necessary one for 
the validity of the  RS assumption.
Since ${\cal G}(s)$ decreases from ${\cal G}(0)$ to zero we 
have also the trivial
consistency condition that the theory developed in the present 
subsection is nontrivial only for $q>0$.

Let us now discuss what happens in the limit $t\gg \tau _J$. We 
rewrite Eq. (\ref{udav5}) taking into account  (\ref{fdtnova-0},
\ref{fdtnova}),
and the concrete value of $r$:
\begin{equation}
\label{udav7}
(1-q)(q^{p-2}-e^{-s}{\cal C}^{p-2}(s)) 
{\cal C}(s)=T\int_0^s d\sdpr e^{- \sdpr }{\cal C}^{p-1}
(\sdpr ){\cal G}(s-\sdpr )
\end{equation}
If we want to investigate the static limit of this equation
we should consider the large-$s$ limit
\begin{equation}
\label{rusalka2}
{\rm lim}_{\tau _J\to \infty} ({\rm lim}_{t\to\infty} C(t))=
\lim_{s\mapsto \infty}{\cal C}(s)= {\cal Q}.
\end{equation}
In this case we have ${\cal Q}=0$, because in the relevant part
 of integral (\ref{udav7})
the function ${\cal G}$ is zero: When $t/\tau_J\to \infty$ every 
correlation will
vanish at large times, and the system goes into paramagnetic phase. 
It is just that statement which
was predicted above starting from some heuristic arguments. 

There is no spin-glass phase if the spins and the coupling constants
have nearly equal characteristic times. It is quite obvious in the light of
the present discussion.

\section{Summary}

This paper is devoted to a glassy system coupled to two heat baths.
In section \ref{sec:2T} we use the adiabatic assumption to generalize 
the usual thermodynamics. Its basic relation (\ref{terra1})
involves the entropies of the spins and the coupling constants, which in the
present approach have the independent and well-defined physical meaning.
The developed theory has a general character, and 
does not depend on concrete details of the considered systems.
After this, in section \ref{sec:stat} this theory is applied to the
mean-field $p$-spin-interaction spherical model, 
extended to have correlated random bonds,
expressed by a finite temperature $T_J$.
In the limit $T_J\to\infty$,
so $n=T/T_J\to 0$,
the usual spin-glass model with totally uncorrelated bonds is recovered. 
As noted recently \cite{allah}, this type of correlations
can make radical changes in the phase
structure. 
In this context, the $p$-spin model 
is the convenient laboratory for investigating
phase transitions, since it belongs to different 
universality classes for $p>2$ and $p=2$.
Indeed, if $n$ is large enough 
there are only first-order phase transitions with positive
latent heat (see Figs. \ref{phase}, 
\ref{phasep2}). This is in the contrast to 
the first-order transitions without
latent heat ($p>2$) or the true second-order transitions
($p=2$), which are more typical for spin glasses
and glasses, and realized in the 
remaining parts of the phase diagram. The 1RSB (replica symmetry breaking)
spin-glass phase can exist as a truly stable phase only for $m<n<1$
(see Eqs. (\ref{prem}, \ref{m})).
Replica symmetry is always restored for 
sufficiently low temperatures and $n>0$.
Notice the differences compared to the SK 
model with infinite-order RSB,
where a finite $n$ mainly modifies the
existing spin-glass phase, 
introducing the lower plateau for the order parameter $q(x)$ 
\cite{kondor}\cite{dotsenko}. 
This plateau grows with $n$, and RSB disappears at some critical value
$n_c$ ($n_c<1$). Nearly the same behavior is introduced by an external 
magnetic field.
In our case such a phase does not exist at all,
and the 1RSB phase with vanishing lower plateau exists even for $n>1$,
but only as a metastable phase.
These distinctions are connected with 
different structures of the phase space.

For all $p\geq 2$ the first-order phase transitions 
are related with an interesting effect: 
In spite of the fact that  the jump of the mean energy $u$ 
at the transition point
is negative (because the latent heat
is positive) the jump of the total entropy  is positive. 
This uncommon property is possible only due to our generalized 
thermodynamical relation (\ref{terra1}), combined with the fact that
the corresponding jumps of $s_{\sigma}$ and $s_J$ have opposite signs: 
$\Delta s_{\sigma}<0$, $\Delta s_J>0$, but the sum $s=s_J+s_{\sigma }$ 
has a positive jump
(see Fig. \ref{fig3}). 
The situation is slightly different for the first-order type phase 
transition without latent heat.
In section \ref{vanish} we have seen that the total entropy
increases continuously in the course of the 
phase transition from the paramagnet to 
1RSB spin-glass (see Fig. \ref{fig00}). 
In this respect an interesting analogy exists
with the process of coarse(fine)-graining (see the discussion
after Eq. (\ref{pun15a})). Further developments of these analogies
will be quite interesting.

There is a close relation between
free energies and saddle-point equations of the finite $n$ RS
case and the $q_0=0$ case of the 1RSB  equations. The
RS solution with nonzero $n$
corresponds to the 1RSB solution with $q_0=0$
\cite{sherington2}. This mechanism is responsible for the 
transition between the 1RSB and RS spin-glass phases. 
The transition is second-order 
with respect to free energy and its derivatives, but is connected 
with a jump of $q_0$ from zero to $q_1$ that ensures 
replica-symmetric behavior (see Figs. \ref{fig1}, \ref{fig2}).
This analogy exists in dynamics also: in our 
time-translation invariant two-temperature
dynamics there is the phenomenon which is the analog
of the generalized FDT in the $n=0$ non-equilibrium dynamics \cite{cug} 
or longtime FDT in the corresponding non-ergodic dynamics
\cite{crisanti2}. Such an effect reflects intrinsic connections 
between systems where the complex
structure of the phase space is self-generated and there
are different time-scales 
for the global relaxation (ergodicity breaking), 
and systems where different components 
have different temperatures and relaxation-scales.

For $n<1$ and $p>2$ there is a difference between phase transitions
predicted by adiabatic statics and adiabatic dynamics.
This well known effect is due to the existence of the whole set of metastable states
with free energies  greater than the
free energy of the pure states predicted by statics. It is
connected with the absence of activated processes on the timescales considered
in the dynamics~\cite{theotwotem}, which enters
due to mean-field (infinite dimension) character of the model
\cite{parisipreprint}. Namely, it is the opposed sequence
of limits: In the true static consideration we
observe times $\to \infty$ before the thermodynamic limit $N\to \infty$,
but if dynamics is investigated by means
of generating functional \cite{dynfunc} the first limit is taken after the
second, which eliminates activated process that need time-scales exponential
in $N$.  In the corresponding finite-dimensional systems a 
smoothening of this effect is expected, where
instead a sharp phase transition 
a near-critical domain of temperatures will take place.
Notice also that for $n=0$ and $p=2$ the RS spin-glass phase
is only marginally stable.
A non-zero $n$ stabilizes the corresponding  fluctuations.

The predictions of the adiabatic statics and dynamics can be compared only
for the relatively short observation times.
The spin-glass phase 
appears at times 
$\ll \tau _J$ (the characteristic time of the coupling constants), but
disappears for the observation times $\gg \tau _J$.
In this limit of long observations 
the coupling constants cannot be viewed as frozen.
This is the non-equilibrium steady state 
without any spin-glass ordering (only a critical slowing down 
of the spin-spin correlation function occurs when $T_J\to 0$).
In this respect it is similar to weak ergodicity breaking occurring
in the non-equilibrium dynamics \cite{cug}.
A generalized fluctuation-dissipation relation 
has been proven, which contains
the  temperature of the couplings instead of the temperature of the spins. 
This relation is also closely connected to the non-equilibrium 
generalization of the FDT \cite{cug,peliti},
where the asymptotic long-time
non-equilibrium state of the $n=0$ $p$-spin spherical model is considered.

\section*{acknowledgments}
A.A. E. is grateful to FOM (The Netherlands) for financial support.
E.Sh. Mamasakhlisov is acknowledged for useful discussions.
D.B. S. thanks Fundacion Andes (Chile) grant c-13413/1, and S. Kobe for 
warm hospitality in Dresden.

\setcounter{equation}{0}

\section*{Appendix A}
\renewcommand{\theequation}{A.\arabic{equation}}

In this appendix we consider how to get the usual Langevin equations of the
spin-glass model with a priori random freezed
coupling constants from (\ref{2}, \ref{3}). 

Eq. (\ref{3}) can be solved
exactly with respect of $J_{\seq }$:
\begin{equation}
\label{aa1}
J_{\seq }(t)=J^{(0)}_{\seq }e^{- (t-t_0)/\tau _J}+A_{\seq }(t)+B_{\seq }(t)
\end{equation}
\begin{equation}
\label{aa2}
A_{\seq }(t)=\frac{1}{\ilt}\int_{t_0}^t d \tdpr  e^{- (t-\tdpr )/\tau _J}
\eta _{\seq }(\tdpr ),
\end{equation}
\begin{equation}
\label{aa3}
B_{\seq }(t)=\frac{1}{\ilt}\int_{t_0}^t d \tdpr e^{-(t-\tdpr )/ \tau _J}
J_N\si _{i_1}(\tdpr )...\si _{i_p}(\tdpr ),
\end{equation}
where $J^{(0)}_{\seq }$ are the initial conditions at the moment $t=t_0$,
$\tau _J=\ilt J_N^2 / v$,
and $J_N^2 =p!J^2/2N^{p-1}$.
The initial factors can be neglected if
$t_0\mapsto  -\infty $, $ | t_0|\gg \tau _J $.
Further we have
\begin{equation}
\label{aa4}
\langle A _{i_1\dots
i_p}(t)A _{j_1\dots j_p}(t^{\prime })\rangle =\frac{T_J J_N^2}{v}
\delta _{i_1\dots
i_p,j_1\dots j_p}e^{- |t-t^{\prime }|/\tau _J},
\end{equation}
Thus if $\tau _J \gg t-t^{\prime }$, then $A_{\seq }$ is a quenched 
Gaussian noise, and if $v\sim T_J$ and $T_J\mapsto \infty$ then
$B_{\seq }$ can be neglected with respect of $A_{\seq }$. So in these limits
$J_{\seq }$ is a quenched Gaussian noise.
Each coupling constant  is independent of the other ones and the spins.

\setcounter{equation}{0}

\section*{Appendix B}
\renewcommand{\theequation}{B.\arabic{equation}} 
In this appendix we discuss the derivation of (\ref{4}) from the 
initial Langevin equations.
We investigate these equations by the dynamical generating functional method
{}~\cite{dynfunc}\cite{thirumalai}.
\begin{equation}
\label{ural1}
1=Z_{{\rm dyn}}=\int \prod_i D\eta _{i}\prod_{[\seq ]}  D\eta _{i_1\dots i_p} 
\exp\left(-\frac{1}{4\Gamma T }\sum _i\int dt
\eta _{i}^2(t)  -\frac{1}{4\ilt T_J }\sum_{[\seq ]}\int dt\eta ^2 _{i_1\dots i_p}(t)\right),
\end{equation}
where $[\seq ]=\lseq $, and 
normalization factors are included in $D\eta _{i} $, $D\eta _{i_1\dots i_p}$.
By means of $Z_{{\rm dyn}}$ we can compute the noise average
of any quantity $A(\{\si \} ,\{J\})$:
\begin{eqnarray}
\label{ural2}
&&\langle A\rangle =
\int D\si DJ\lef [\frac{\de \eta }{\de \si}\rig ]\lef [\frac{\de \eta }{\de
J}\rig ]A(\{\si \} ,\{J\})
\exp[ -\frac{1}{4\Gamma T }\sum_i\int dt \left
( \Gamma \pdt \si _{i}+r\si _{i}+\frac{\partial \H}{\partial \si _{i}} \right
)^2\nonumber \\
&&-\frac{1}{4\ilt T_J}\sum_{[\seq ]} \int dt \left  ( \ilt \pdt
J_{i_1\dots i_p}
+\frac{\partial  \H}{\partial J_{i_1\dots i_p}}\right  )^2]
\end{eqnarray}
where $[\de \eta /\de \si ]$, $[\de \eta /\de J]$ are the corresponding
functional Jacobians. After a simple transformation we have
\begin{eqnarray}
\label{ural22}
&&
Z_{{\rm dyn}}=\int D\si DJ D\hat{\si}D\hat{J}
\exp \left ( -\Gamma T\sum_i\int dt\hat{\si }^2_i +  i\sum_i\int dt\hat{\si
}_i(t)
\left ( \Gamma \pdt \si _{i}+r\si _{i}+\frac{\partial
\H}{\partial \si _{i}}\right )\right )\nonumber \\
&&\exp\left ( \sum_{[\seq ]}\left [-\ilt T_J\int dt \hat{J}^2_{\seq }(t)
+i \int dt\hat{J}_{\seq }(t)\left (\ilt \pdt J_{i_1\dots i_p}
+\frac{\partial \H}{\partial J_{i_1\dots i_p}} \right ) \right ] \right )
\nonumber \\
&& \exp (V_{\eta \si }+V_{\eta J})
\end{eqnarray}
Here the last exponent is the contribution of the functional Jacobians; we
shall not write this expression explicitly
because it is not relevant in the mean-field approximation.

We want to derive equations for spin dependent functions,
therefore in eq. (\ref{ural1}) we can take Gaussian integrations by $\{J\}$,
$\{\hat{J}\}$.
The result is
\begin{eqnarray}
\label{ural25}
&& Z_{{\rm dyn}}=\int D\si D\hat {\si }\exp  [ -\Gamma T \sum_i\int dt
\hat{\si}_i(t)+i\sum_i\int dt \hat{\si }_i(t)(\Gamma \pdt +r) \si _t \nonumber \\
&& -\frac{i J_N^2}{\tilt }\sum _{[\seq ]}\int dtdt^{\pr }\phi (t-t^{\pr
})a_{\seq }(\si , \hat{\si },t^{\pr })
b_{\seq }(\si ,t) \nonumber \\
&& -\frac{T_J J^2_N}{2v }\sum _{[\seq ]}\int dtdt^{\pr }
k(t-t^{\pr })a_{\seq }(\si , \hat{\si },t^{\pr })
a_{\seq }(\si ,\hat{\si },t)  ]\nonumber \\  
&& \exp (V_{\eta \si }+V_{\eta J}),
\end{eqnarray}
where
\begin{equation}
\label{ural24}
a_{\seq}(\si ,\hat{\si },t)=\sum_{s=1}^p\si _{i_1}(t)...
\si _{i_{s-1}}(t)\hat{\si }_{i_s}(t)\si _{i_{s+1}}(t)
...\si _{i_p}(t), \,
b_{\seq }(\si ,t)=\si _{i_1}(t)...\si _{i_p}(t),
\end{equation}
\begin{equation}
\label{ural222}
\phi (t-t^{\pr})=\theta (t-t^{\pr})e^{- (t-t^{\pr })/\tau _J}, \,
k(t-t^{\pr})=e^{- |t-t^{\pr }|/\tau _J}, \, \tau_J=\Gamma_J/v
\end{equation}
By the standard mean-field procedure  \cite{thirumalai} we introduce the
following order parameters:
\begin{eqnarray}
\label{ural26}
&& Q_1(t,t^{\pr })=\lan i\hat{\si }(t)i\hat{\si }(t^{\pr })\ran ,\nonumber \\
&&Q_2(t,t^{\pr })=\lan \si (t)\si (t^{\pr })\ran ,\nonumber \\
&& Q_3(t,t^{\pr })=\lan \si (t)i\hat{\si }(t^{\pr })\ran , \nonumber \\
&& Q_4(t,t^{\pr })=\lan \si (t^{\pr })i\hat{\si }(t)\ran ,
\end{eqnarray}
and the corresponding Lagrange factors $\la _s(t,t^{\pr })$, $s=1,..,4$. 
In this scheme $Q_2$ is the correlation function, $Q_3$, $Q_4$ 
are the susceptibilities; $Q_1$ (``field-field'' correlation function)
should be taken zero by reasons of causality \cite{thirumalai}.
Now we have
\begin{eqnarray}
\label{ural27}
&& Z_{{\rm dyn}}=\int\prod_{s=1}^4\frac{D\la _sDQ_s}{2\pi i}\exp \left(N\Omega (\la _s, Q_s)+
N\int D\si D\hat{\si } e^{{\cal L}(\si ,\hat{\si } )} \right
)\nonumber \\
&&\Omega (\la _s, Q_s)=-\int dt dt^{\pr }\sum_{s=1}^4\la _s(t,t^{\pr
})Q_s(t,t^{\pr })
+\frac{pJ^2}{2\tilt }\int dt dt^{\pr }\phi (t-t^{\pr })
Q_3(t,t^{\pr })Q^{p-1}_2(t,t^{\pr })\nonumber \\
&& +\frac{pT_J J^2}{4v}\int dt dt^{\pr }\phi (t-t^{\pr })
[Q_1(t,t^{\pr })Q^{p-1}_2(t,t^{\pr })+(p-1)Q_3(t,t^{\pr })Q_4(t,t^{\pr
})Q^{p-2}_2(t,t^{\pr })],
\end{eqnarray}
where
\begin{eqnarray}
\label{ural28}
&&{\cal L}(\si ,\hat{\si })=-\Gamma T \sum_i\int dt \hat{\si}_i(t)-i\sum_i\int dt
\hat{\si }_i(t)(\Gamma \pdt +r) \si _t
\nonumber \\
&&+\la _1(t,t^{\pr })i\hat{\si } (t)i\hat{\si } (t^{\pr })
+\la _2(t,t^{\pr })\si (t)\si (t^{\pr }) \nonumber\\
&& +\la _3(t,t^{\pr })i\si  (t)\hat{\si }  (t^{\pr })+\la _4(t,t^{\pr
})i\hat{\si } (t)\si  (t^{\pr })
\end{eqnarray}
By variational methods we obtain
\begin{eqnarray}
\label{ural29}
&&\la _1(t,t^{\pr })=\frac{pT_J J^2}{4v}k(t-t^{\pr })Q^{p-1}_2(t,t^{\pr })
\nonumber \\
&&\la _2(t,t^{\pr })=\frac{p(p-1)J^2}{2\tilt }\phi (t-t^{\pr })
Q_3(t,t^{\pr })Q^{p-2}_2(t,t^{\pr })\nonumber \\
&&+\frac{pT_J J^2}{4v}k(t-t^{\pr })
((p-1)Q_1(t,t^{\pr })Q^{p-2}_2(t,t^{\pr })+(p-1)(p-2)Q_3(t,t^{\pr })Q_4(t,t^{\pr })
Q^{p-3}_2(t,t^{\pr}))
\nonumber \\
&&\la _3(t,t^{\pr })=\frac{pJ^2}{2\tilt }
\phi (t-t^{\pr })Q^{p-1}_2(t,t^{\pr })
+(p-1)\frac{pT_J J^2}{4v}k(t-t^{\pr })Q_4(t,t^{\pr }) Q^{p-2}_2(t,t^{\pr })
\nonumber \\
&& \la _4(t,t^{\pr })=(p-1)\frac{pT_J J^2}{4v}k(t-t^{\pr })Q_3(t,t^{\pr })
Q^{p-2}_2(t,t^{\pr }),
\end{eqnarray}
$\la _2$ can be adsorbed in the Jacobians.
These results should be substituted to (\ref{ural28}): the effective dynamics
of a spin is determined by
spins motion at the environment of the spin, and by motion of the coupling
constants. After this lengthy calculation we arrive at (\ref{4}).


\end{document}